\documentclass[aip,jcp,amssymb,amsmath,reprint]{revtex4-2}
\usepackage{graphicx}
\usepackage{amsthm}
\usepackage{comment}

\def\be{\begin{equation}}
\def\ee{\end{equation}}
\def\ber{\begin{eqnarray}}
\def\eer{\end{eqnarray}}
\def\bern{\begin{eqnarray*}}
\def\eern{\end{eqnarray*}}

\def\rv{\mathbf{r}}

\def\Rv{\mathbf{R}}

\def\uv{\mathbf{u}}

\def\0v{\mathbf{0}}
\def\1v{\mathbf{1}}
\def\2v{\mathbf{2}}
\def\3v{\mathbf{3}}

\def\pa{\partial}

\DeclareMathAlphabet\mathbfcal{OMS}{cmsy}{b}{n}

\begin{document}

\title{High-frequency limit of spectroscopy}

\author{Vladimir~U.~Nazarov}

\email{nazarov.vu@mipt.ru}
\affiliation{Moscow Institute of Physics and Technology (National Research University), Dolgoprudny, Russian Federation}
\affiliation{Fritz Haber Research Center for Molecular Dynamics and Institute of Chemistry, Hebrew University of Jerusalem, Jerusalem, Israel}

%\email{vladimir.nazarov@mail.huji.ac.il}
%\affiliation{Fritz Haber Research Center for Molecular Dynamics and Institute of Chemistry, Hebrew University of Jerusalem, Jerusalem, Israel}

\author{Roi Baer}
\email{roi.baer@huji.ac.il}
\affiliation{Fritz Haber Research Center for Molecular Dynamics and Institute of Chemistry, Hebrew University of Jerusalem, Jerusalem, Israel}

\begin{abstract}
We consider an arbitrary quantum mechanical system, initially in its ground-state, exposed to a time-dependent electromagnetic pulse with a carrier frequency $\omega_0$ and a slowly varying envelope of finite duration. 
By working out a solution to the time-dependent Schr\"odinger equation in the high-$\omega_0$ limit, we find that, to the leading order in $\omega_0^{-1}$, a perfect self-cancellation of the system's linear response occurs as the pulse switches off. Surprisingly, the system's observables are, nonetheless, describable in terms of a combination of its linear density response function and nonlinear functions of the electric field.
An analysis of jellium slab and jellium sphere models reveals a very high surface sensitivity of the considered setup, producing a richer excitation spectrum than accessible within the conventional linear response regime. 
On this basis, we propose a new spectroscopic technique, 
which we provisionally name the Nonlinear High-Frequency Pulsed Spectroscopy (NLHFPS).
Combining the advantages of the extraordinary surface sensitivity, the absence of constraints by the traditional dipole selection rules, and the clarity of theoretical interpretation utilizing the linear response time-dependent density functional theory, 
NLHFPS has a potential to evolve into a powerful characterization method for nanoscience and nanotechnology. 
\end{abstract}

%The spectral transitions turn out to not depend on the electronic dipole operator rules but other observables.

\maketitle

\section{Introduction}

In optical spectroscopy, systems of interest are exposed to light, and their response allows us to explore their structure and composition. A significant part of spectroscopy involves linear effects, such as when light is absorbed or scattered off material targets, allowing their imaging and characterization, teaching us almost solely about dipole-allowed transitions. Nonlinear spectroscopy methods go beyond this limitation, studying otherwise hidden or dark changes applicable to a large variety of systems and processes.%
\cite{Mukamel}  Examples of nonlinear spectroscopy include the second-order harmonic generation (SHG) approach, used to study interfaces and adsorbed molecules and serves as high-resolution optical microscopy in biological systems,\cite{roke_nonlinear_2012} multiphoton excitation fluorescence (MPEF), as well as various Raman scattering methods.\cite{johansson_nonlinear_2018}  

The use of nonlinear spectroscopies, especially in surface and nano-sciences, is growing due to their high interfacial sensitivity. However, results in nonlinear spectroscopies are often  challenging to interpret since 
their description involves much more sophisticated theoretical techniques as compared to their linear counterparts.
\cite{Mukamel2006}

This article studies the high-frequency limit of the electronic response, singling out a pathway which leads to a major simplification in the description of nonlinear spectroscopies, as long as the observables are analysed {\em after the field acting on a system dies out}. We find that the nonlinear behaviour of the system observables is expressible in terms of the \emph{linear electron density response function}, the latter occurring on the time-scale of the pulse's enveloping shape.  
By this, we present an approach to the problem of the nonlinear electronic response in the case of high-frequency pulses, which turns out no more theoretically and computationally demanding than the solution of the conventional linear response problem. Specifically, the well-developed methods of the linear response time-dependent density functional theory\cite{Gross-85}
(TDDFT)  can be readily invoked, expanding the reach of the latter to the realm of the nonlinear physics.

We validate our theory numerically using the exactly solvable hydrogen atom system propagating under a time-dependent field. Then we consider applications to nano-films and nano-dots, which demonstrate the power of the proposed method by revealing the modes in the excitation spectra of these systems, latent when probed within the linear regime. 
Finally, we present an example of molecular spectroscopy showing dipole-forbidden transitions.

\section{Formalism}
\label{formalism}

We consider a many-electron system subject to the time-dependent (TD) modulated periodic potential. We are concerned with solving the Schr\"odinger equation 
(in the following, atomic units are used unless indicated otherwise) 
\begin{equation}
i \frac{\pa \Psi(t)}{\pa t}= \left[\hat{H}_0 + (\cos\omega_0 t) \hat{W}(t) \right] \Psi(t),
\label{SE}
\end{equation}
where the unperturbed Hamiltonian is
\begin{equation}
\hat{H}_0= \sum\limits_{i=1}^N  \left[ -\frac{1}{2} \nabla_i^2 +v_{ext}(\rv_i)\right] + 
\frac{1}{2} \sum\limits_{i\ne j}^N \frac{1}{|\rv_i-\rv_j|},
\label{H0}
\end{equation}
$N$ and $v_{ext}(\rv)$ being the number of electrons and the external (electron-nuclear Coulomb) potential, respectively,
and the harmonic perturbation is enveloped with the potential
\begin{equation}
\hat{W}(t)=\sum\limits_{i=1}^N W(\rv_i,t).
\end{equation}
For simplicity, we assume that the time-dependence in the pulse potential $W(\rv,t)$ factorizes, i.e.,
\begin{equation}
W(\rv,t)=C(t) W(\rv),
\label{Fact}
\end{equation}
where $C(t)$ is the pulse envelope and $W(\rv)$ determines the coordinate dependence of the  potential, 
although, extensions to more general forms of the potential are straightforward.

Our principal result, the proof of which is postponed until Appendix~\ref{AD} and the Supplemental Material, is an expression for the probability amplitude to find, after the end of the pulse, the system in its excited state $\Psi_{\alpha\ne 0}$
\begin{equation}
\begin{split}
\langle \Psi_{\alpha\ne 0}|\Psi(t>T)\rangle &= 
\frac{\pi  \widetilde{C^2}(E_\alpha  -  E_0 ) }{2 i\omega_0^n} e^{-i E_\alpha t}\\
&\times \int \langle \Psi_\alpha | \hat{n}(\rv) |\Psi_0 \rangle  F_n(\rv)  d\rv .
\end{split}
\label{princ}
\end{equation}
In Eq.~(\ref{princ}), $E_\alpha$ are the eigenenergies of the system, 
$\widetilde{C^2}(\omega)$ is the Fourier transform of the \emph{square} of the envelope function
\begin{equation}
\widetilde{C^2}(\omega) = \frac{1}{2 \pi}  \int e^{i \omega t} C^2(t) d t, 
\end{equation}
$\hat{n}(\rv)=\sum_{i=1}^N \delta(\rv_i-\rv)$ is the electron density operator, 
and $n=4$ and $2$, in the case of the uniform applied electric field 
\begin{equation}
W(\rv)=- \mathbfcal{E}_0 \cdot \rv,
\label{dipW}
\end{equation}
and all other cases, respectively. Corresponding $F_n(\rv)$ are
\begin{align}
&F_4(\rv)= -[ \mathbfcal{E}_0 \cdot \nabla]^2 v_{ext}(\rv),
\label{F}\\
&F_2(\rv)= [\nabla W(\rv)]^2.
\label{Fnd}
\end{align}
Finally, Eq.~(\ref{princ}) holds to the leading non-vanishing order  $\omega_0^{-n}$ in each of the cases.
Further developments (see Appendix~\ref{AD}) show, that the time-dependent oscillations in the electron density after the end of the pulse are given by
\begin{equation}
\delta n(\rv,t>T)  \! = \!  \frac{ 1}{i \omega_0^n}  
\!  \int \! e^{-i \omega t}
 \widetilde{C^2}(\omega) {\rm Im} \chi(\rv,\rv',\omega)  
 F_n(\rv') d\rv' d \omega,
\label{nassy}
\end{equation}
where $\chi(\rv,\rv',\omega)$ is the linear density response function of the interacting electron system.%
\footnote{The RHS of Eq.~(\ref{nassy}) can easily be seen real, the presence of the imaginary unity in the denominator notwithstanding, which is due to the oddness of ${\rm Im}\, \chi(\rv,\rv',\omega)$ in $\omega$ and to the fact that 
${\widetilde{C^2}^*(-\omega)=\widetilde{C^2}(\omega)}$, the latter due to the realness of $C(t)$.}

Furthermore, to the leading order in $\omega_0^{-1}$, we find for the total energy absorbed by the system during the pulse action
\begin{equation}
\Delta E \! = \! - \frac{\pi }{4 \omega_0^{2 n}} 
 \! \! \int \! \!  \omega |\widetilde{C^2}(\omega)|^2  
F_n(\rv)
{\rm Im} \chi(\rv,\rv',\omega) 
F_n(\rv')
 d \omega d\rv d\rv'.
\label{DE}
\end{equation}

Clearly, the  case of the uniform electric field ($n=4$) is relevant to the problem of the illumination  by light.
Although, strictly speaking, the latter should be described with the  transverse vector potential $A_z(t-x/c)$, the usual practice is, neglecting the retardation, to reduce the problem to that with the homogeneous $A_z(t)$ and then, 
by the gauge transformation, to the equivalent problem with the scalar potential (\ref{dipW}). \cite{Landau-71} 
%\cite{Eberly-93,Barash-99,Vorobeichik-99,Balanarayan-13}.
Apart from the lower bound on the frequency, inherent to our high-frequency asymptotic theory, $\omega_0 \gg \omega_\text{low}$, the neglect of the retardation imposes a standard upper bound  $\omega_0 \ll \omega_\text{high}=c/d$, where $c$ is the velocity of light, and $d$ is the size of the system.
Another case, $n=2$, is relevant to processes with  the excitation by  {\em longitudinal} fields, such, e.g., as with moving charges.%
\footnote{The potential $\phi_{ext}(\rv,t)=Z/|\rv-\Rv(t)|$ of an ion of the charge $Z$ moving along the trajectory $\Rv(t)$ 
corresponds to the non-uniform externally applied field, except for $|\rv-\Rv(t)|\gg |\rv|$ } 
This is promising for the construction of TDDFT of the {\em stopping power} of matter for fast ions beyond the adiabatic approximation for the exchange-correlation potential, which theory now exists in the low-velocity limit only. \cite{Nazarov-05,Nazarov-07}

Importantly, in Eqs.~(\ref{nassy}) and (\ref{DE}) we witness a hybridization of linear and quadratic response quantities:  the \emph{linear} density-density response function is multiplied by the \emph{quadratic} frequency envelop $\widetilde{C^2}(\omega)$. In the illustrative calculations below, we will see that such hybridization leads to interesting effects. 

In the field of the  light-matter interactions, the application of the acceleration-frame method of Kramers and Henneberger (KH)  \cite{Kramers-56,Henneberger-68} has led to a great many advancements in the theory.%
\cite{Eberly-93,Barash-99,Vorobeichik-99,Baer-09,Eckardt-15,Ben-Asher-20}
Instructively, our formulas above can be re-derived in an alternative way using the KH method, as it is shown in Appendix \ref{AKH}. However, this is possible to do in the case of the uniform field only ($n=4$), since this case is inherent within  
the KH formalism.

\section{Results}

\subsection{Hydrogen atom}  
\begin{figure} [h!] 
\includegraphics[width= \columnwidth, trim= 0 0 0 0, clip=true]{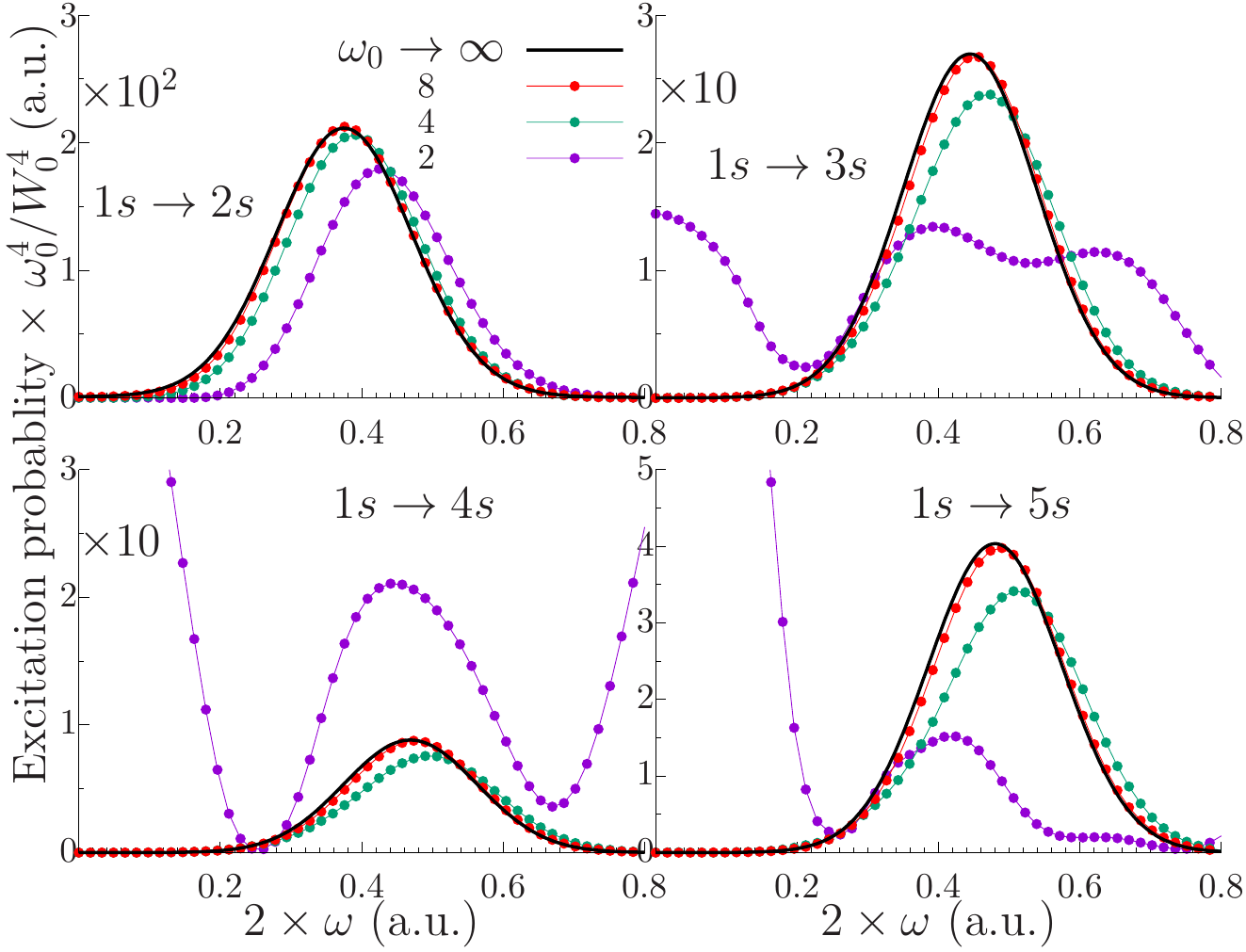}
\caption{\label{ndhyd}
Excitation probability [the modulus squared of Eq.~(\ref{princ})] upon the end of the pulse of Eq.~(\ref{ndpulse}), from the ground-state of the hydrogen atom to a number of its excited $s$-states. The solid black line is the asymptotic limit of Eq.~(\ref{matndipH}). Spectra at finite frequencies are obtained by the numerical propagation of the TD Schr\"{o}dinger equation (\ref{SE}). The parameters of the pulse used were $\sigma=15$ a.u. and $W_0=0.125$ a.u.}
\end{figure}

We now investigate how the high-frequency limit is approached as the frequency increases by calculating a precisely solvable system, namely, the hydrogen atom. 
First, assuming an atom, initially in its ground state, is subjected to the doubly modulated Gaussian pulse with a spherically symmetric quadrupole potential
\begin{equation}
W(\rv,t) \cos\omega_0 t=W_0 r^2 e^{-(t/\sigma)^2} \cos \omega t \cos\omega_0 t,
\label{ndpulse}
\end{equation}
we numerically time-propagate the Scr\"{o}dinger equation (\ref{SE}). 
In the pulse (\ref{ndpulse}), the carrier frequency $\omega_0$ serves to set the scene for the high-frequency regime, while the second frequency $\omega$ couples the pulse to the excitations in the system.
Upon the end of the pulse, we look at the populations of the excited states, plot them in Fig.~\ref{ndhyd} versus the enveloping function frequency $\omega$ (the {\it second} frequency), and compare  with the asymptotic limit. 
The latter, according to Eqs.~(\ref{princ}), (\ref{Fnd}), and (\ref{ndpulse}) is given by
\begin{equation}
 \langle \phi_{n, s} | \phi(t>T)\rangle \! = \!
\frac{2 \pi W_0^2}{ i\omega_0^2} e^{-i \epsilon_n t}
\widetilde{C^2}(\epsilon_n-\epsilon_1 ) 
 \langle \phi_{n,s}(r) | r^2 |\phi_{1,s}(r) \rangle,
\label{matndipH}
\end{equation}
where $\phi_{n,s}(r)$ are the hydrogenic $s$-orbitals and $\epsilon_n$ are the corresponding eigenenergies,
and we have restricted the comparison to the transitions to the $s$-states only.
We note that the spherically symmetric quadrupole potential (\ref{ndpulse}) is purely model one, which we use to demonstrate the convergence of the numerical solution of the Schr\"{o}dinger equation to the asymptotic solution (\ref{princ}) for a non-uniform field ($n=2$).

Similarly, in the  case of the uniform field ($n=4$), we propagate the system under the  potential
\footnote{While the potential of Eq.~(\ref{ndpulse}) is purely model, we note that the regime of Eq.~(\ref{dpulse}) can be realized by superimposing two lasers' beams.}
\begin{equation}W(\rv,t) \cos \omega_0 t=-\mathcal{E}_0 z e^{-(t/\sigma)^2} \cos \omega t \cos \omega_0 t.
\label{dpulse}
\end{equation}
For the hydrogen atom 
\begin{equation}
\begin{split}
F_4(\rv)  &=  \mathcal{E}_0^2 \frac{\pa^2}{\pa z^2} \frac{1}{r} =  - \mathcal{E}_0^2 \\
& \times \left[\frac{ (4\pi)^{3/2}}{3} \delta(\rv) Y_{0 0}(\theta,\phi) \! + \!
 \sqrt{\frac{\pi}{5}} \frac{4}{r^3} Y_{2 0}(\theta,\phi)\right],
\end{split}
\end{equation}
where $Y_{l m}(\theta,\phi)$ are  spherical harmonics. Evidently, only transitions from the ground state to s- and d-states are possible, which have the following amplitudes
\begin{align}
\begin{split}
&\langle \phi_{n>1, s}|\phi(t>T)\rangle = -\frac{\pi \mathcal{E}^2_0 \widetilde{C^2}(\epsilon_n  -  \epsilon_1)  }{2 i \omega_0^4}  
e^{-i \epsilon_n t} \\
&\times \frac{ 4\pi}{3} \phi_{n,s}(0) \phi_{1,s}(0),
\end{split} \label{p0}\\
\begin{split}
&\langle \phi_{n>2, d}|\phi(t>T)\rangle = \frac{\pi \mathcal{E}^2_0 \widetilde{C^2}(\epsilon_n  -  \epsilon_1)  }{2 i \omega_0^4}  
e^{-i \epsilon_n t}\\
&\times 4 \sqrt{\frac{\pi}{5}} \int\limits_0^\infty \frac{1}{r} \phi_{n,d}(r) \phi_{1,s}(r) dr.
\end{split}
\label{p2}
\end{align}
where, in Eq.~(\ref{p0}), we can further simplify with account of  $\phi_{n,s}(0)=2/n^{3/2}$. \cite{Landau-81}
\begin{figure} [h!] 
\includegraphics[width= \columnwidth, trim= 8 0 0 0, clip=true]{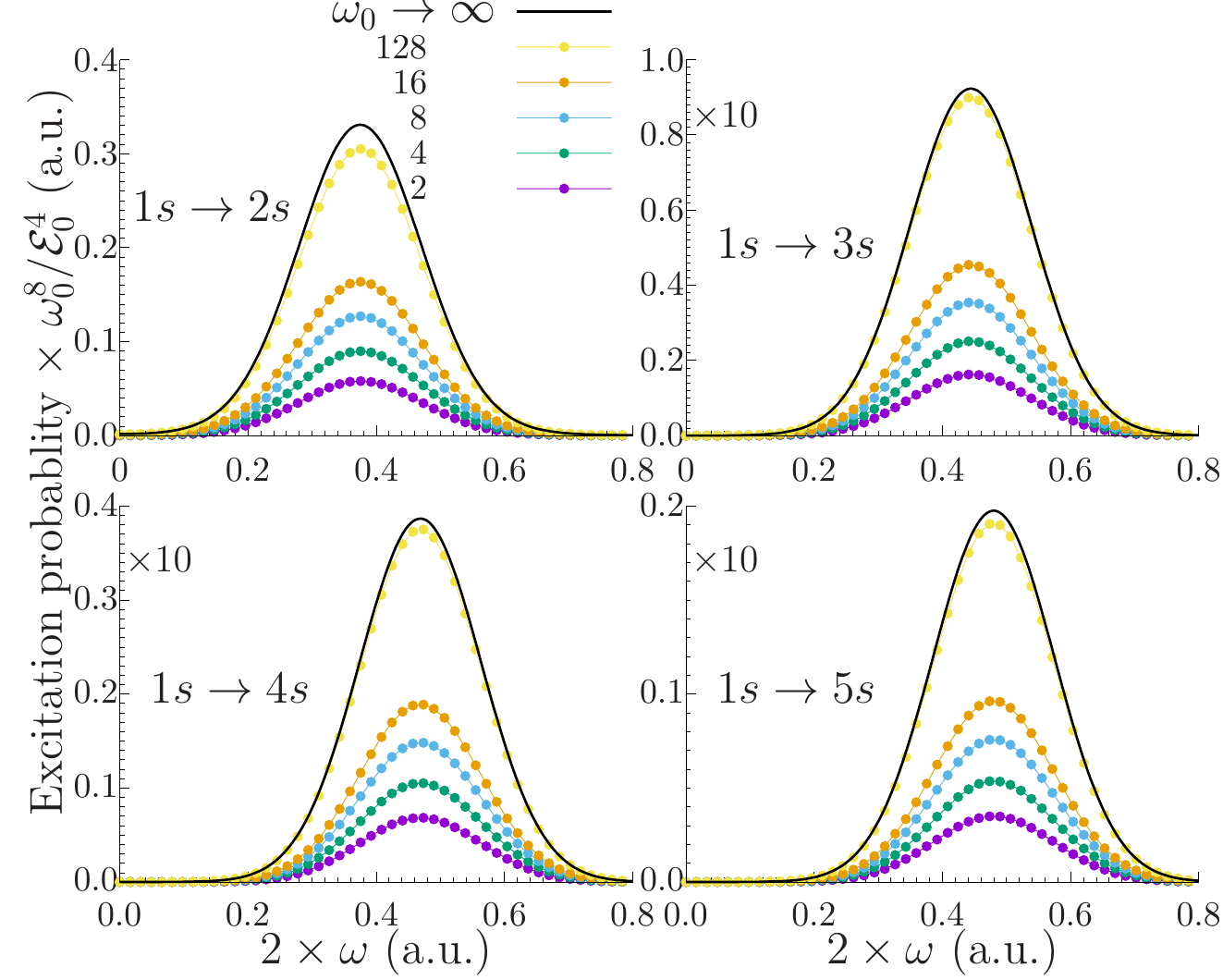}
\caption{\label{dhyd}
Excitation probability, upon the end of the pulse of Eq.~(\ref{dpulse}), from the ground-state of the hydrogen atom to some of its excited $s$-states. The solid black line is the asymptotic limit of Eq.~(\ref{p0}). Spectra at finite frequencies are obtained by the numerical propagation of the TD Schr\"{o}dinger equation. The parameters of the pulse used were $\sigma=15$ a.u. and $\mathcal{E}_0=0.125$ a.u.}
\end{figure}

\begin{figure} [h!] 
\includegraphics[width= \columnwidth, trim= 2 0 0 0, clip=true]{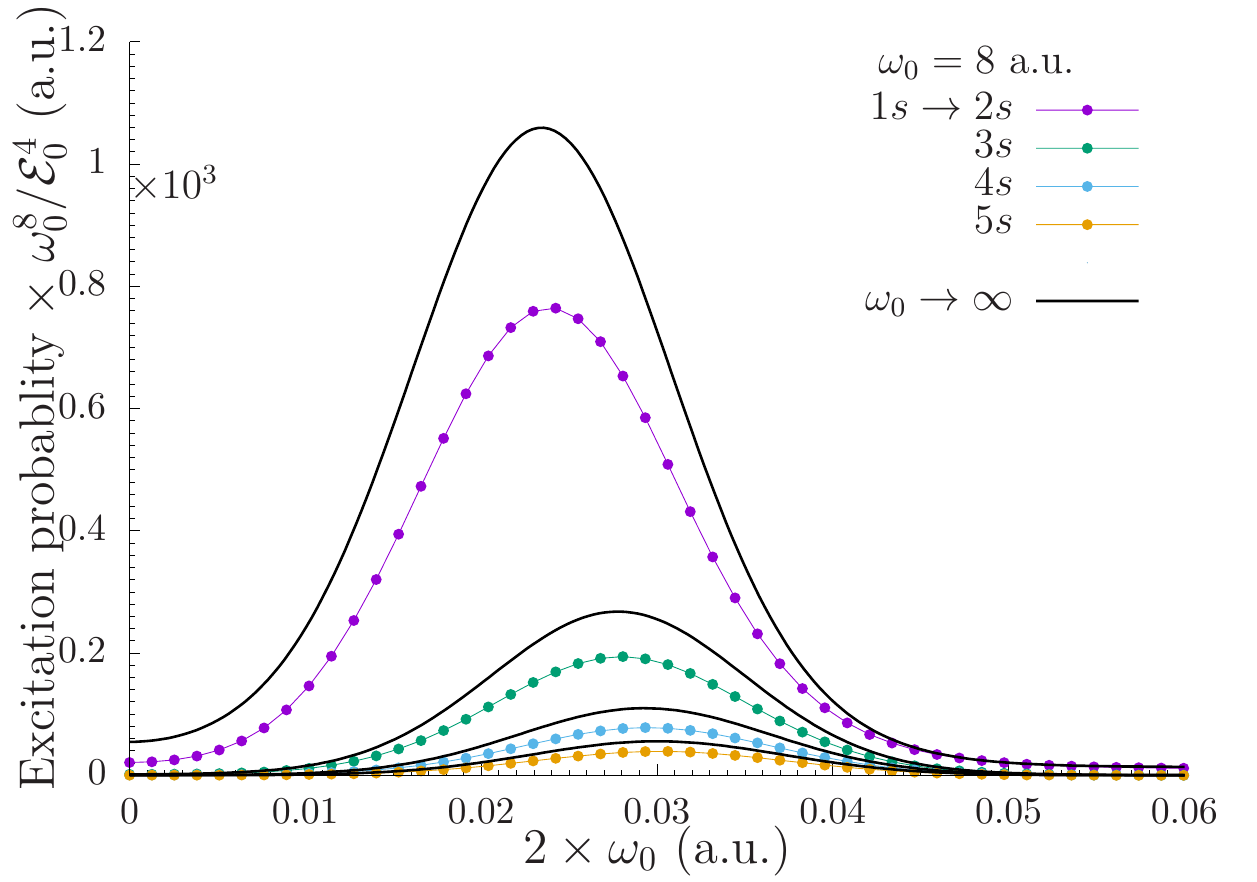}
\caption{\label{Z025w8}
Similar to Fig.~\ref{dhyd}, but for the fictitious hydrogenic atom of the nuclear charge $Z=0.25$ and with $\sigma=200$ a.u.}
\end{figure}

Figures \ref{ndhyd} and \ref{dhyd}  demonstrate the convergence, with the growth of $\omega_0$, 
of the excitation processes' outcome to their $\omega_0\to \infty$ limits of Eqs.~(\ref{princ}), for the cases of the quadrupole and dipole exciting potentials, respectively. 
Remarkably, in the quadrupole (Fig.~\ref{ndhyd}) and  the dipole (Fig.~\ref{dhyd}) cases, the asymptotic  regime is approached
in very different ways: 
in the former case, the peaks' positions and shape change dramatically with the frequency growth, 
while in the latter, the amplitude of the peaks varies monotonously only.
In the dipole case,  the convergence, with respect to peaks' amplitudes,  is very slow, and it is not reached at practically achievable values of $\omega_0$. %
\footnote{Apart from the experimental unachievability of the upper values of $\omega_0$ in Fig.~\ref{dhyd}, at those frequencies results become unphysical because of the retardation effects, as discussed in Sec.~\ref{formalism}. 
The rationale for our including these high frequencies is to confirm that, although slow,  the convergence takes place nonetheless.
}
At the same time, the excitation energies (peaks' positions), even at moderate values of $\omega_0$, are very well reproduced by the asymptotic theory.
We point out and emphasize that, 
while the asymptotic limit holds for an arbitrary system, the speed of the convergence is system-dependent.
This is confirmed by Fig.~\ref{Z025w8} with the use of the fictitious system of the hydrogenic atom with the nuclear charge of $Z=0.25$.
Since the asymptotic theory is expected to be the more accurate the larger is $\omega_0$ compared to the characteristic excitation energies in a system, in the $Z=0.25$ case we observe a much faster convergence compared to the $Z=1$.
For peaks in Fig.~~\ref{Z025w8} to remain resolved, a large width of the pulse $\sigma=200$ a.u. was chosen in the calculation with $Z=0.25$.
Further particulars of the solution of the TD Schr\"{o}dinger equation and the issues of the convergence to the asymptotic limit are presented in Appendix \ref{AConv}.

\begin{figure} [h!] 
\includegraphics[width= \columnwidth, trim= 40 0 10 0, clip=true]{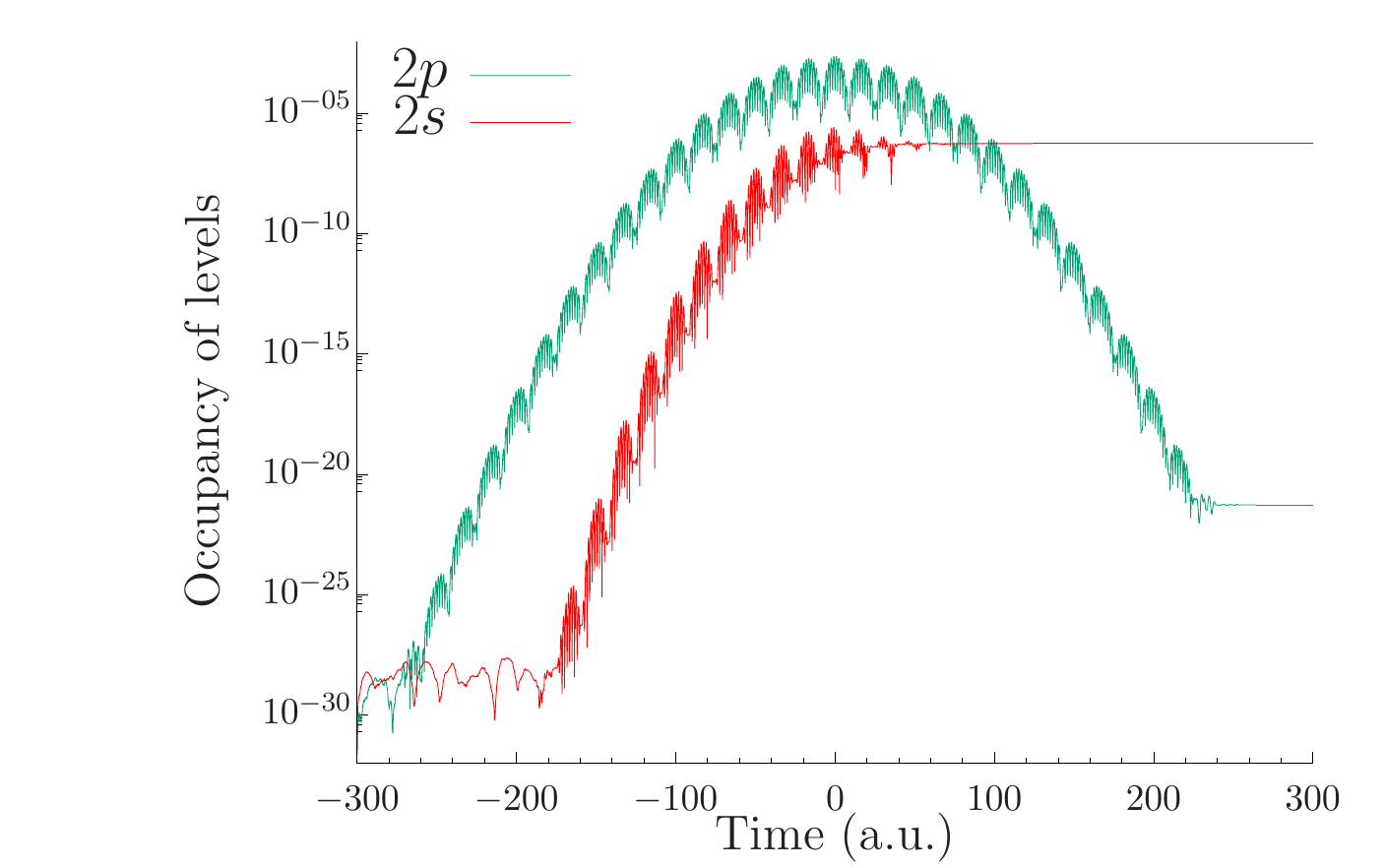}
\caption{\label{prophyd}
Evolution of the populations of orbitals in H atom during the action of the pulse of Eq.~(\ref{dpulse}). 
Parameters used were $\omega_0=2$ a.u., $\omega=(\epsilon_{2}-\epsilon_{1})/2=0.1875$ a.u., $\sigma=50$ a.u., and $\mathcal{E}_0=0.125$ a.u.}
\end{figure}

It is highly instructive to follow the excitation process in time, from the pulse beginning to its end,
in order to understand how the system reaches its final state. 
As can be seen from the derivation [e.g., Eq.~(S.15) of the Supplemental Material], the linear response does contribute to the pumping during the pulse action, but it passes a cycle from increasing to decreasing the population of excited states, with the zero net result. On the contrary, the quadratic response does not completely reverse itself, which results in the residual occupancies of the excited states upon the pulse’s end.  
In Fig.~\ref{prophyd}, we plot the time-evolution of the population numbers of the $2s$- and $2p$- orbitals of H atom under the action of the pulse of Eq.~(\ref{dpulse}). 
We observe the principal difference between the change of the occupancies of the $s$- and $p$- levels: while the latter gets much more (approximately three orders of magnitude) populated in the middle of the pulse duration, it gives the electron away upon the pulse end. At the same time, the former keeps the accepted electron with a finite probability. This type of behaviour is characteristic of spherically symmetric systems in the high-frequency regime, which is in agreement with our asymptotic theory. This is the linear response that dominates the $s\to p$ transition at the time of the pulse duration, which is gone upon the pulse's extinction. 
In particular, we conclude that the usual dipole selection rules do not hold in this process.

At this point we note that, while TDDFT  of the electronic response in the high-frequency limit was studied in Ref.~\onlinecite{Nazarov-10},
it is important to emphasize the principal difference between the physical situation considered in that reference  and
in the present paper. 
Ref.~\onlinecite{Nazarov-10} deals with the response to the monochromatic field, thus considering a {\em continuous wave}.
In that regime, the linear response persists in the  high-frequency limit and it is, usually, prevailing.
On the contrary, here we consider the excitation by a {\em pulse of finite duration}, the carrier frequency of which is asymptotically high. We focus on the behaviour of a system after the end of the pulse, in which case we find the total suppression of the linear response, while the nonlinear one is describable in terms of the linear response TDDFT.

\begin{figure} [h!] 
\includegraphics[width= \columnwidth, trim= 55 0 0 0, clip=true]{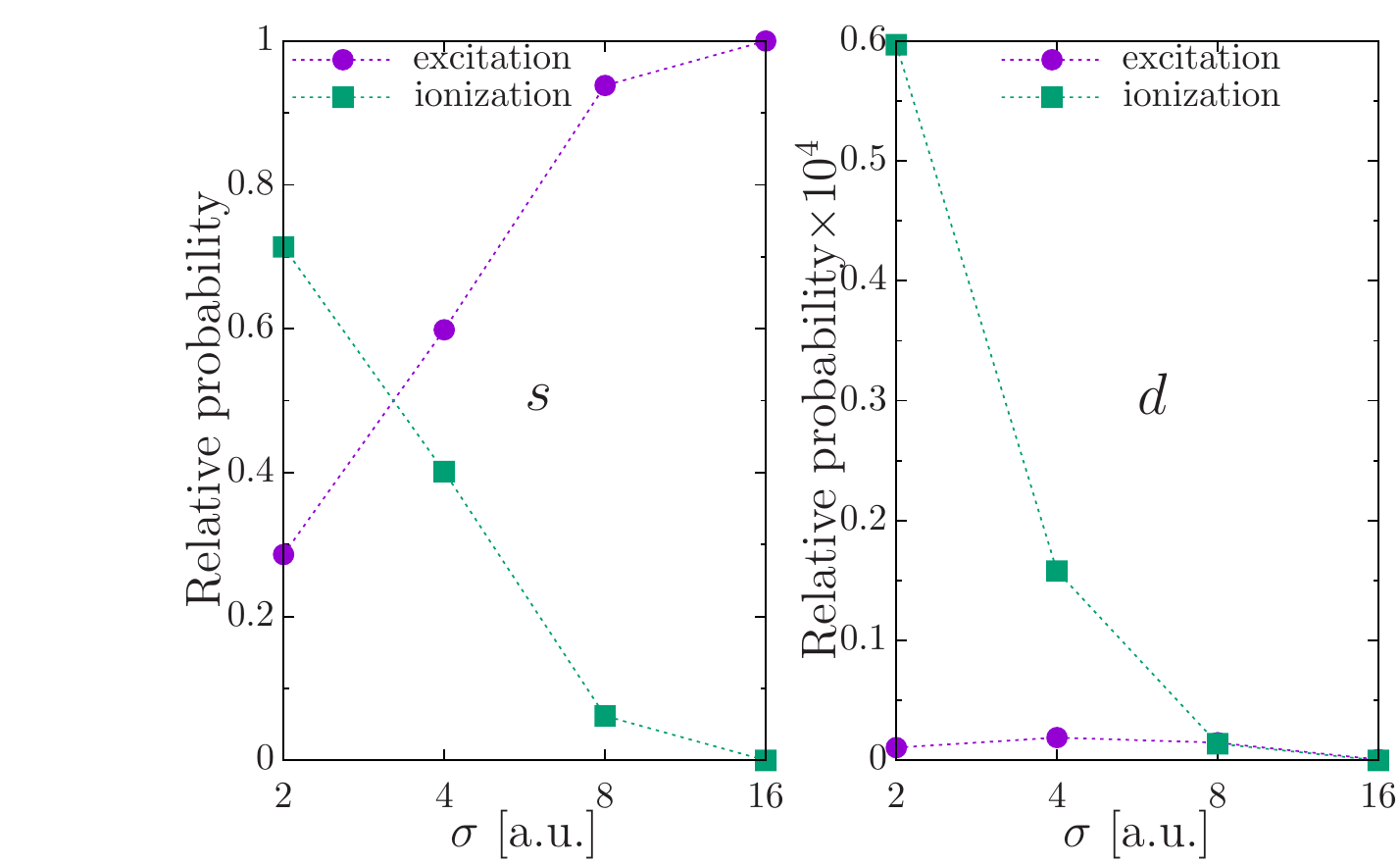}
\caption{\label{hyexio}
Probability of the excitation and ionization of hydrogen atom, initially in its ground state, to $s$- (left) and $d$- (right) states,
relative to the total excitation plus ionization probability, plotted versus the pulse width $\sigma$.
The pulse shape is purely Gaussian $C(t)=e^{-(t/\sigma)^2}$.
}
\end{figure}

With the use of Eqs.~(\ref{p0}) and (\ref{p2}), in Fig.~\ref{hyexio} we compare the excitation and ionization processes' probabilities for the hydrogen atom initially in its ground-state and exposed to the Gaussian pulse.
We conclude that the ionization is dominant for short pulses, in which case a sudden impact strips off electron,
while, for longer pulses,  transitions to excited bound states become preferential.
We also note that transitions to the $d$-states play insignificant role compared to those to the $s$-states.

\subsection{Jellium slab}
\begin{figure} [h!] 
\includegraphics[width= \columnwidth, trim= 70 0 0 0, clip=true]{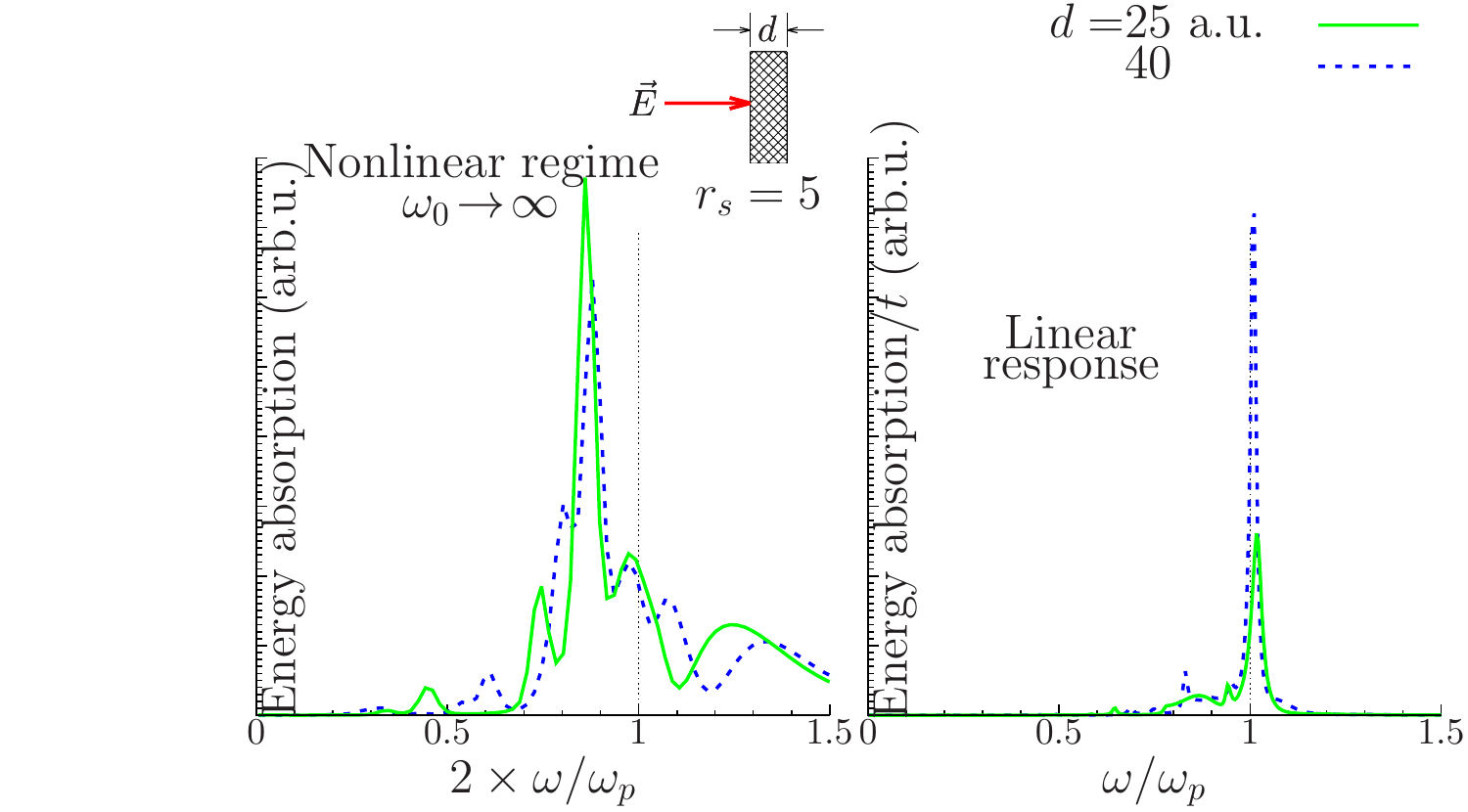}
\caption{\label{lK}
Jellium slabs. Left: absorption from the pulse of Eq.~(\ref{dpulse}) ($\sigma= 500$ a.u.) at asymptotically large frequency $\omega_0$ as a function of $ 2\omega$, as obtained through  Eq.~(\ref{DE}). Right: absorption per unit time from the monochromatic field of the frequency $\omega$ in the linear response regime.
Two slabs of the thicknesses $d=25$ and $40$ a.u. and the density parameter $r_s=5$  are considered.
$x$-axes are scaled to the bulk plasma energy $\omega_p=4.2$ eV. Parameters used correspond to the jellium model of solid potassium.
The inset shows the slab geometry and an arrow indicates the direction of the electric field vector, while the laser pulse moves parallel to the slab's surfaces.}
\end{figure}

\begin{figure} [h!] 
\includegraphics[width= \columnwidth, trim= 70 0 0 0, clip=true]{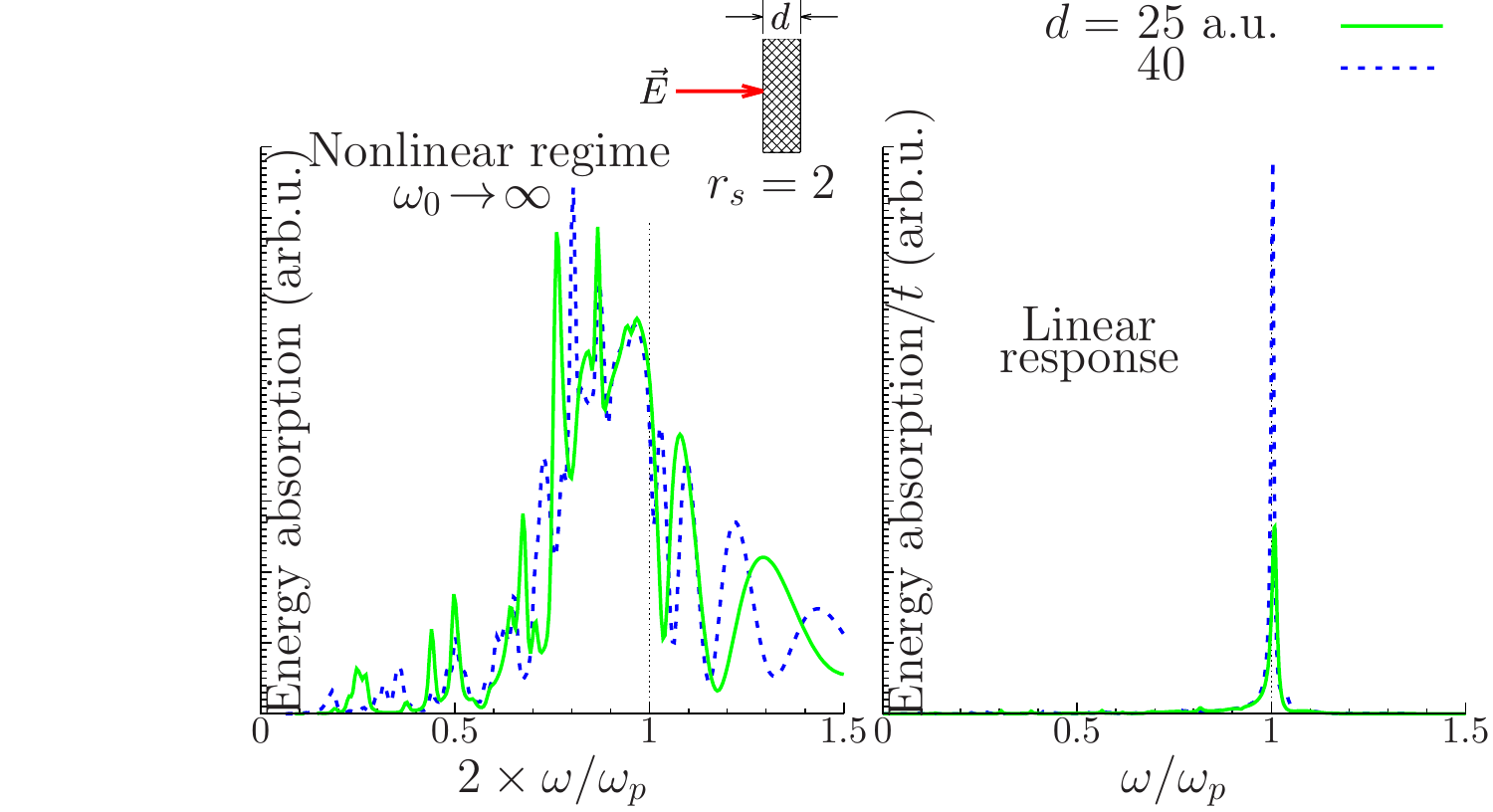}
\caption{\label{loss}
Same as Fig.~\ref{lK}, but for slabs of the density parameter $r_s=2$
and the corresponding  bulk plasma energy $\omega_p=16.7$ eV (jellium model of solid aluminum).
}
\end{figure}

We proceed by considering a slab of the thickness $d$ with  the positive constant background charge density 
$n_+= (\frac{4}{3} \pi r_s^3)^{-1}$, where $r_s$ is the 3D density parameter.
Within the Kohn-Sham (KS) density-functional theory (DFT) \cite{Kohn-65} and using the local density approximation (LDA), 
we calculate the ground-state KS band structure and electron density.
To this system, we apply the doubly modulated dipole pulse of Eq.~(\ref{dpulse}),
and we use our theory to determine the total energy absorption in the slab in the high carrier frequency regime.
The problem being one-dimensional, the operator in Eq.~(\ref{F}) reduces to the Laplacian, and we have by virtue of the Poisson law
\begin{equation}
F_4(z)= - 4 \pi  n_+(z)=
-4\pi n_+\Theta \! \left(\frac{d}{2}-|z|\right),
\label{Fs}
\end{equation}
where $\Theta(x)$ is the Heaviside's step-function.
Resulting absorption spectra, obtained by  Eq.~(\ref{DE}) with the use of the adiabatic time-dependent LDA (ATDLDA) 
in the construction of $\chi(\rv,\rv',\omega)$, \cite{Gross-85} are presented in Figs.~\ref{lK} and \ref{loss},
for $r_s=5$ and $2$, corresponding to the jellium model of the metallic potassium and aluminum, respectively. 
The following observations are made:
(i) Similar to the case of the hydrogen atom, due to the integration with $\widetilde{C^2}(\omega)$ in  Eq.~(\ref{DE}) and due to the form of the pulse (\ref{dpulse}), 
spectra in the left panels of Figs.~\ref{lK} and \ref{loss} as functions of $\omega$ are governed by SHG and, accordingly, peaks' positions scale to half the frequencies of the corresponding excitations;
(ii) In the linear regime (right panels in Figs.~ \ref{lK} and \ref{loss}),  spectra are dominated by the bulk plasmon (BP) peak, the intensity of which crucially depends on the share of the bulk, i.e., the slab thickness $d$.
On the contrary, the nonlinear spectra in the high-frequency regime (left panels in Figs.~\ref{lK} and \ref{loss}) 
weakly depend on $d$, suggesting that the surface excitations dominate them.
The prevalence of the surface response can be  understood by noting that 
$\int \chi(\rv,\rv',\omega) d \rv'=0$ (no reaction to a constant potential)
and, therefore, both the deep  interior and exterior of the slab, by Eq.~(\ref{Fs}), do not contribute  appreciably to the integral of Eq.~(\ref{DE}).

Notably, in the left panel of Fig.~\ref{lK} we observe a strong peak with the maximum at $2 \omega \approx 0.88 \omega_p$. The counterpart of this peak in the linear response regime (right panel of Fig.~\ref{lK}) is positioned at $\omega\approx 0.83 \omega_p$, and it is known as the multipole surface plasmon (MP). \cite{Tsuei-90-0} Because of the BP suppression, MP is very prominent in the left panel of this figure, which makes the high-frequency nonlinear technique an ideal tool to study this otherwise subtle type of excitation. 
It is instructive to note that $F_4(z)$ of Eq.~(\ref{F})
provides, effectively, the {\em impact} mode of the complementary linear response problem, \cite{Liebsch} which  is known to be favourable for MP excitation. \cite{Nazarov-99} 
In Fig.~\ref{loss} ($r_s=2$), left panel, we also see a prominent broad peak at $2 \omega$ below the BP frequency, while MP is not discernible in the linear response spectrum in the right panel.
 We, therefore, conclude that the corresponding excitation exists at the surface of metallic aluminum, and the high-frequency nonlinear technique provides a unique way to detect it. At the same time, the traditional method of electron energy loss spectroscopy (EELS) does not possess sufficient sensitivity. \cite{Tsuei-90-0}
The oscillating structures at $2\omega>\omega_p$ in Figs.~\ref{lK} and on both sides from $\omega_p$ in Fig.~\ref{loss} differ for different slab thicknesses, and they can, therefore, be attributed to the interference effect between the two surfaces of the slabs.
Finally, the absence of the conventional (dipole) surface plasmon (SP) peak at $\omega_s=\omega_p/\sqrt{2}$  is due to the strictly normal to the surface direction of the exciting field ($q_\|=0$), in which case the amplitude of the SP vanishes.

\begin{figure} [h!] 
\includegraphics[width= \columnwidth, trim= 90 0 70 0, clip=true]{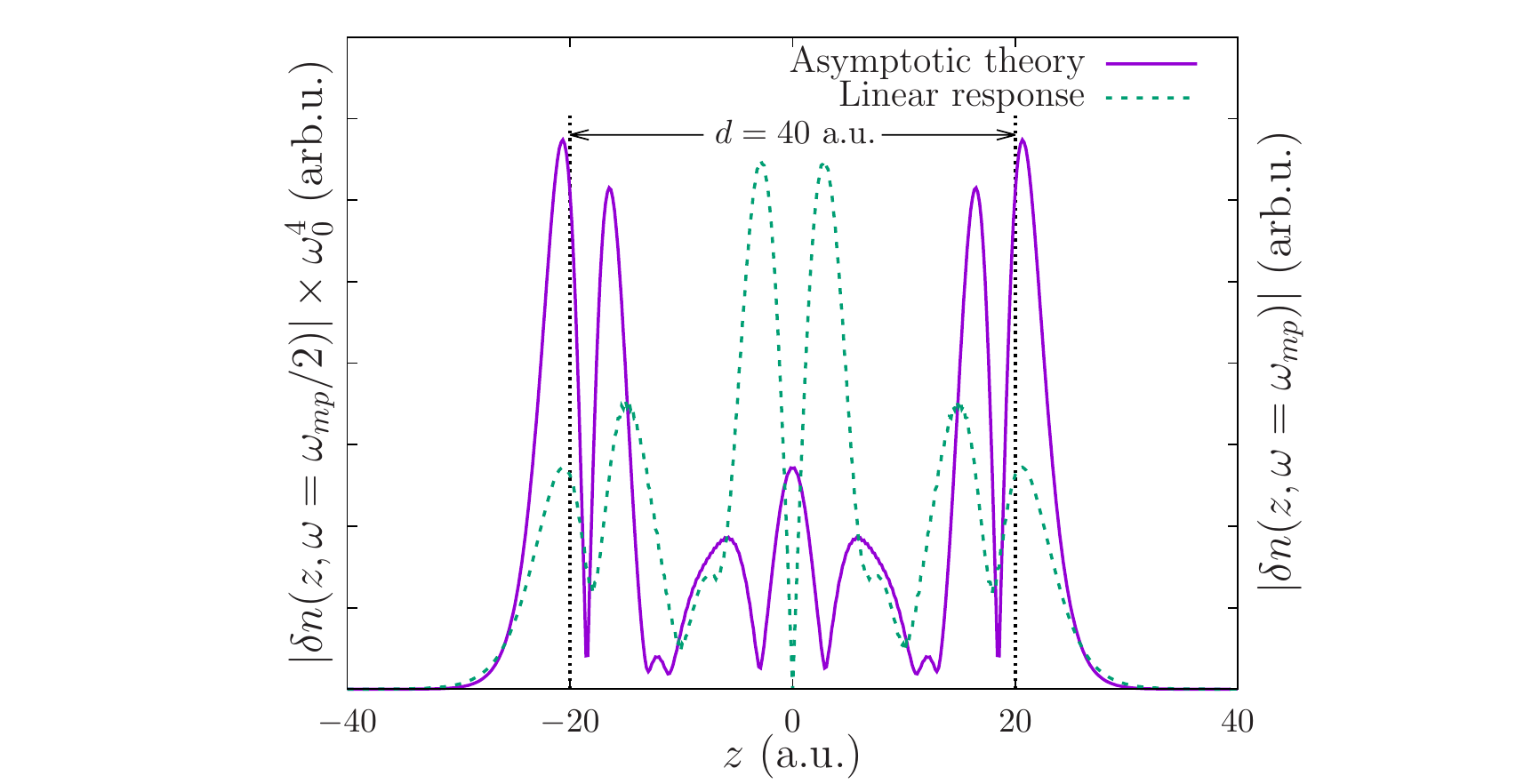}
\caption{\label{n1}
Fourier transform of the density oscillation  [Eq.~(\ref{nassy})] in the $\omega_0\to\infty$ asymptotic regime 
(solid curve against the left $y$-axis) and its linear response counterpart (dashed curve against the right $y$-axis), 
with the frequency $\omega$ set to $\omega_{mp}/2$ and $\omega_{mp}$, respectively [{\it cf}. Fig.~(\ref{lK})].
Vertical straight lines indicate positions of the slab's surfaces.
Parameters of the calculation are those of Fig.~\ref{lK}.
}
\end{figure}

To quantitatively verify the above picture, in Fig.~\ref{n1} we plot the Fourier transform of the  density oscillation in the asymptotic regime [Eq.~(\ref{nassy})] and compare it with the linear response density oscillation.
Clearly, in the former case, the oscillation is mainly confined to the vicinity of the surfaces of the slab being largely suppressed in the interior.
On the contrary, in the linear response regime, oscillations predominantly occur in the bulk of the slab.

\subsection{Jellium sphere}
In contrast to a slab, for a sphere, the second derivative in the RHS of Eq.~(\ref{F}) does not reduce to Laplacian
and, consequently, $F_4(\rv)$ is not given by the positive background density only. Instead, we have
\begin{equation}
\begin{split}
&F_4(\rv) =
\frac{2 \sqrt{4 \pi } n_+}{3}  \times \\
&\left[\frac{2}{\sqrt{5}} \frac{R^3}{r^3} \Theta(r-R) Y_{20}(\theta,\phi)-\Theta(R-r) Y_{00}(\theta,\phi) \right],
\end{split} 
\label{Fsp}
\end{equation}
where $R$ is the radius of the rigid positive-charge background. Due to the symmetry, the density-response function $\chi(\rv,\rv',\omega)$ splits in angular momentum into $\chi_{l m}(r,r',\omega)$, the latter acting separately on each  harmonic of the externally applied potential. The problem becoming one-dimensional again, we calculate $\chi_{0 0}$ and $\chi_{2 0}$, apply them to Eq.~(\ref{Fsp}), and plug the result into Eq.~(\ref{DE}). We consider the same form of the doubly modulated pulse of Eqs.~(\ref{dpulse}) as previously.
\begin{figure} [h!] 
\includegraphics[width= \columnwidth, trim= 70 0 0 0, clip=true]{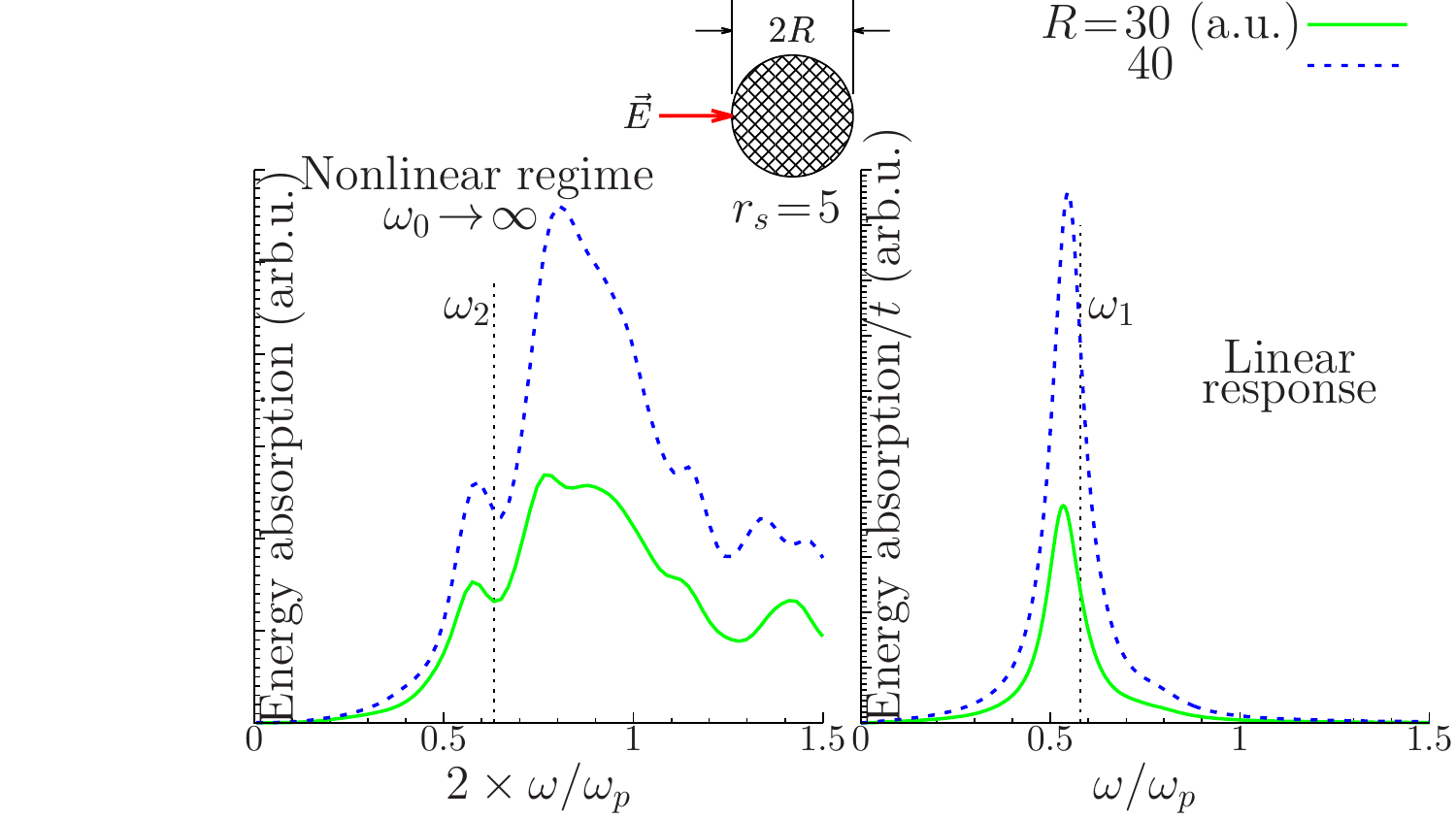}
\caption{\label{sprs5}
Jellium spheres. Left: absorption from the pulse of Eq.~(\ref{dpulse})  ($\sigma= 500$ a.u.) at asymptotically large frequency $\omega_0$ as a function of $ 2\omega$, as obtained through  Eq.~(\ref{DE}). Right: absorption per unit time from the monochromatic field of the frequency $\omega_0$ in the linear response regime.
Vertical lines show positions of classical Mie plasmons $\omega_l$. 
Two spheres of the radii $R=30$ and $40$ a.u. and the density parameter $r_s=5$  are considered.
%$x$-axes are scaled to the bulk plasma energy $\omega_p=4.2$ eV. 
}
\end{figure}

In Fig.~\ref{sprs5}, results of calculations for two spheres, with radii $R=30$ and $40$ a.u., and the density parameter $r_s=5$, are presented, for the nonlinear $\omega_0\to\infty$ and the linear-response regimes, in the left and right panels, respectively. Within the classical electrodynamics, a sphere of the Drude metal supports an infinite series of Mie plasmons $\omega_l=\sqrt{l/(2 l+1)} \omega_p, \ l=1,2,\dots$ .\cite{Bohren-98}
%\begin{equation}
%\omega_l=\sqrt{\frac{l}{2 l+1}} \omega_p, \ l=1,2,\dots.
%\label{Mie}
%\end{equation}
In the monochromatic linear-response (right panel of Fig.~\ref{sprs5}), we observe the $p$-mode only of this series, red-shifted by the quantum size effect. 

According to Eq.~(\ref{Fsp}), energy absorption in the nonlinear $\omega_0\to\infty$ regime (left panel of Fig.~\ref{sprs5}) originates from the superposition of the $s$- and $d$-modes. 
As plotted versus the second modulation frequency $\omega$,
it reveals a rich spectrum of the underlying excitations.
The leftmost feature near $0.57 \omega_p$ comes from the $d$-mode Mie plasmon $\omega_2$, red-shifted in the quantum calculation.
The broad dominating peak with the maximum near $0.80 \omega_p$ does not have an analog within the classical electrodynamics, and, similar to the multipole plasmon modes in the case of a slab, it becomes accessible with the use of the high-$\omega_0$ nonlinear regime. A signature of the bulk plasmon on the right shoulder of this peak can also be observed, indicating the possibility of the direct recognition of the constituents of  nano-particles by their bulk plasmon frequencies 
$\omega_p$ with the use of laser pulses. The latter is, obviously, impossible in the linear-response regime. We also note structures above $\omega_p$, which are due to the (dressed) single-particle excitations affected by the quantum interference.

\begin{figure*}  
\includegraphics[width= \textwidth]{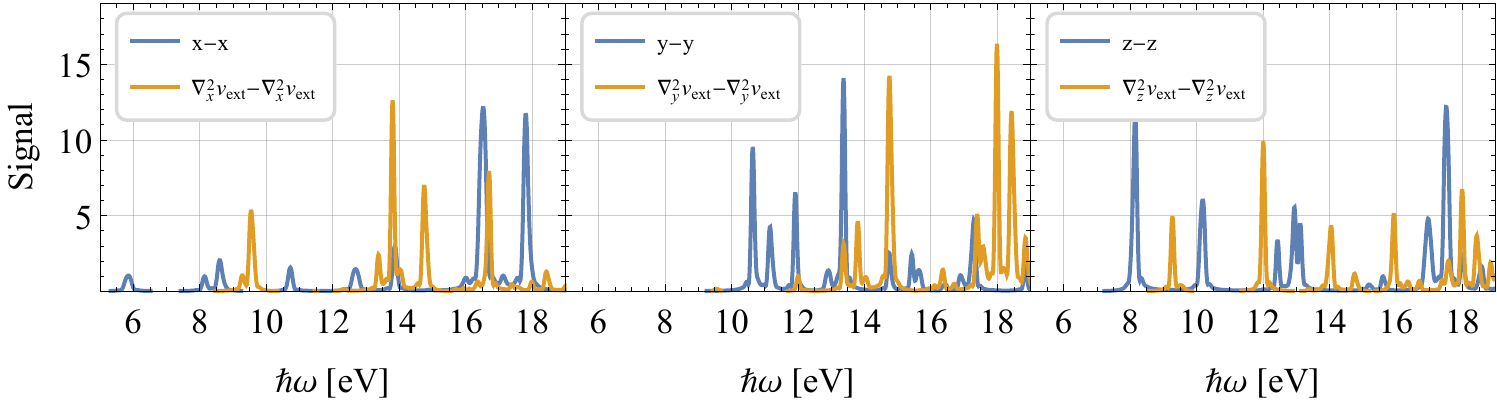}
\caption{\label{ethylene}
Comparison of the standard linear-response energy absorption spectrum of the ethylene molecule to that of the high-frequency response in the dipole approximation (Eq. \ref{DE}). $x-x$ refers to the linear response auto-correlation function with the electric field along the $x$-axis. At the same time, $\nabla_x v_{ext}-\nabla_x v_{ext}$ stands for the auto-correlation function in the high-frequency nonlinear regime, and the similarly for two other directions.
}
\end{figure*}

Finally, we consider molecular electronic spectroscopy. Referring back to the above-discussed very short pulse spectroscopy, in the dipole interaction case, we used time-dependent local density approximation calculations to produce linear response estimates of the high-frequency energy absorption (using Eq. \ref{DE}) and, in Fig.~\ref{ethylene}, compare to standard low-frequency energy absorption for the ethylene molecule. 
The spectra' differences in the two regimes are due to the dipole versus $F_4$ selection rules,
emphasizing the high-frequency spectroscopy's aptitude to probe the excitations forbidden in the linear regime. See Appendix  \ref{AR} for details concerning this calculation.

\section{Discussion and conclusions}

We have considered  excitation of a quantum-mechanical system by an externally applied electric field of 
high-frequency $\omega_0$ and finite duration in time.
After the end of the pulse,  the state of the system being a superposition of the eigenstates of the unperturbed Hamiltonian, the expansion of the corresponding transition amplitudes  in the power series in 
$\omega_0^{-1}$  has been performed, with the leading terms found of the order $\omega_0^{-4}$ for the uniform applied field (dipole case) and of $\omega_0^{-2}$, otherwise.

We have demonstrated that, to the leading order in the inverse frequency, the quadratic, rather than the linear, response determines the excitation process. Nonetheless, we have also shown that all the information necessary to describe this nonlinear excitation regime is contained in the linear density response function of the system under consideration. The problem has been thus reduced to that of the linear response time-dependent density functional theory, for which practical methods of solution, at various levels of accuracy and sophistication,  are well established.

Further, we have found that a specific pulse shape, 
modulation by the second (low) frequency can be advantageous as a probe, delivering spectra of excitations in the nonlinear response regime.
In our illustrative applications, to the jellium model nano-films and nano-dots, plasmonic modes undetectable or challenging for the detection by the linear optical spectroscopy or electron energy-loss spectroscopy have been discerned. We point out that the high carrier frequency is out of resonance, and its only role is to set the scene for probing the system with the second frequency, the twice of the latter being in resonance with the system's excitations.

Based on our findings, we propose a spectroscopic technique, which we provisionally name
the Nonlinear High-Frequency Pulsed Spectroscopy. 
Our results show that NLHFPS, i.e., exposing an explored system to a finite-duration high-frequency electric field with low-frequency modulation, allows for an efficient nonlinear spectroscopic probe of modes inaccessible or hardly accessible by other techniques. A significant asset of the novel method is its ease of interpretation, enabling a detailed comparison between experiment and theory. This benefit stems from the results' direct dependence on the target material's density-density response function. As demonstrated here, NLHFPS can uncover rich and profound physical phenomena hidden from more conventional methods.

\section{Supplementary Material}

Supplementary Material contains detailed derivation of Eq.~(\ref{AL}) of the Appendix \ref{AD}, which is too lengthy to be placed in the main text or appendices.

\acknowledgments

V.U.N. acknowledges the support of the
Russian Foundation for Basic Research and the Ministry of Science and Technology of Taiwan (Grant no. 21-52-52007).
R.B. wishes to acknowledge the support of the German-Israel Foundation (Grant no. GIF-I-26-303.2-2018).

\

\noindent
The authors declare no conflicts of interest.

\

\noindent
The data that support the findings of this study are available from the authors upon reasonable request.

\appendix

\section{Derivation of Eqs.~(\ref{princ})-(\ref{DE})}
\label{AD}
In the interaction representation 
\begin{align}
&\tilde{\Psi}(t) = e^{i \hat{H}_0 t} \Psi(t), \\
&\tilde{\hat{W}}(t)  = e^{i \hat{H}_0 t} \hat{W}(t) e^{-i \hat{H}_0 t},
\label{dPsi}
\end{align}
the problem of the solution of Eq.~(\ref{SE}) turns  into that for the  equation
\begin{equation}
\frac{\pa}{\pa t} \tilde{ \Psi}(t) = \frac{1}{i}
(\cos\omega_0 t) \tilde{\hat{W}}(t)
 \tilde{\Psi}(t),
 \label{diff}
\end{equation}
or for the equivalent integral equation
\begin{equation}
\tilde{ \Psi}(t) = \Psi_0 + \frac{1}{i} \int\limits_{-\infty}^t 
(\cos\omega_0 t') \tilde{\hat{W}}(t')
 \tilde{\Psi}(t')  d t',
 \label{inteq}
\end{equation}
where by $\Psi_\alpha$  we denote the  set of eigenfunctions of the Hamiltonian (\ref{H0}), we assume that $W(\rv,-\infty)=0$, and the system is initially in its ground-state $\Psi_0$.

Performing several consecutive integrations by parts in Eq.~(\ref{inteq}), assuming 
the pulse to be of  finite duration [$\hat{W}(\rv,+\infty)=0$] and $\omega_0$ to be large, we obtain, after keeping the terms up to $\omega_0^{-4}$ only
\begin{widetext}
\begin{equation}
\begin{split}
&\tilde{ \Psi}(+\infty) = \Psi_0  + 
\frac{1}{4 \omega_0^2} \! \int\limits_{-\infty}^\infty
\! \left[ \frac{\pa \tilde{\hat{W}}(t)}{\pa t}, \tilde{\hat{W}}(t) \right] \! \Psi_0 d t 
 - 
\frac{1}{4 \omega_0^4} \! \int\limits_{-\infty}^\infty
\! \left[ \frac{\pa^3 \tilde{\hat{W}}(t)}{\pa {t}^3}, \tilde{\hat{W}}(t) \right] \! \Psi_0 d t
 + 
\\
&\frac{1}{16 \omega_0^4} \! \! \int\limits_{-\infty}^\infty 
\! \left[ \frac{\pa \tilde{\hat{W}}(t)}{\pa t}, \tilde{\hat{W}}(t)  \right]
 \int\limits_{-\infty}^{t} \! \left[ \frac{\pa \tilde{\hat{W}}(t')}{\pa t'}, \tilde{\hat{W}}(t') \right]
\! \Psi_0 d t' d t
\! - \! 
\frac{1}{16 \omega_0^4} \! \! \int\limits_{-\infty}^\infty \!
\left[ \frac{\pa \tilde{\hat{W}}(t)}{\pa t} ,\tilde{\hat{W}}^3(t) \right]  \Psi_0 d t
\! + \! \frac{3}{64 \omega_0^4} \!  \! \int\limits_{-\infty}^\infty 
\left[ \frac{\pa \tilde{\hat{W}}^2(t)}{\pa t}, \tilde{\hat{W}}^2(t)\right]  \Psi_0 d t .
\end{split}
\label{AL}
\end{equation}
A lengthy derivation of Eq.~(\ref{AL}) is given in full in the Supplementary Material.
\footnote{Arriving at final concise Eqs. (\ref{princ})-(\ref{DE}) has required very lengthy derivations. To rule out a possibility of error, we have repeated the derivation several times. We have also verified results by an independent method using the Kramers-Henneberger’s acceleration frame (Appendix \ref{AKH}). Additionally, after the manual derivation, we composed a computer algebra code (in Mathematica) for consecutive integrations by parts in Eq. (\ref{inteq}), which produced exactly the same results.}
The commutators in Eq.~(\ref{AL}) can be expanded as 
\begin{equation}
\left[ \frac{\pa \tilde{\hat{W}}(t)}{\pa t}, \tilde{\hat{W}}(t) \right]=
i e^{i \hat{H}_0 t}  \left[ \left[\hat{H}_0,\hat{W}(t)\right] ,\hat{W}(t)\right] e^{-i \hat{H}_0 t}, 
\label{comm1}
\end{equation}
\begin{equation}
\begin{split}
&\left [\frac{\pa^3 \tilde{\hat{W}}(t)}{\pa t^3},\tilde{\hat{W}}(t)\right]= 
e^{i \hat{H}_0 t}  \left[
-i  \left[\hat{H}_0,  \left[ \hat{H}_0,  \left[\hat{H}_0,\hat{W}(t) \right] 
\right]  \right]   
- 3 \left[ \hat{H}_0,  \left[\hat{H}_0,\frac{\pa \hat{W}(t)}{\pa t} \right] 
\right] +
3 i \left[\hat{H}_0, \frac{\pa^2 \hat{W}(t)}{\pa t^2} 
\right] ,\hat{W}(t)\right]  e^{-i \hat{H}_0 t} ,
\end{split}
\label{d3}
\end{equation}
\end{widetext}
\begin{equation}
\left[ \frac{\pa \tilde{\hat{W}}(t)}{\pa t}, \tilde{\hat{W}}^3(t) \right]=
i e^{i \hat{H}_0 t}  \left[ \left[\hat{H}_0,\hat{W}(t)\right] ,\hat{W}^3(t)\right] e^{-i \hat{H}_0 t} ,
\label{comm3}
\end{equation}
\begin{equation}
\left[ \frac{\pa \tilde{\hat{W}}^2(t)}{\pa t}, \tilde{\hat{W}}^2(t) \right]=
i e^{i \hat{H}_0 t}  \left[ \left[\hat{H}_0,\hat{W}^2(t)\right] ,\hat{W}^2(t)\right] e^{-i \hat{H}_0 t}. 
\label{comm4}
\end{equation}

\subsection{Non-uniform field case}
We evaluate the commutator  (\ref{comm1}) to
\begin{equation}
\begin{split}
\left[ \left[\hat{H}_0,\hat{W}(t)\right] ,\hat{W}(t)\right] 
= - \int [\nabla W(\rv,t)]^2 \hat{n}(\rv) d\rv,
\label{ndip0}
\end{split}
\end{equation} 
If the RHS of Eq.~(\ref{ndip0}) is not zero, then the substitution of Eq.~(\ref{ndip0}) into Eq.~(\ref{AL}), keeping only the leading term of the order $\omega_0^{-2}$,
yields
\begin{equation}
\begin{split}
&\langle \Psi_{\alpha\ne 0}|\tilde{\Psi}(t>T)\rangle = 
\frac{1}{4 i\omega_0^2} \times \\
& \int  
e^{i(E_\alpha-E_0) t} \langle \Psi_\alpha | \hat{n}(\rv) |\Psi_0 \rangle [\nabla W(\rv,t)]^2  d\rv  d t.
\end{split}
\label{ndip}
\end{equation}
If, furthermore, the factorization of Eq.~(\ref{Fact}) holds,
then we arrive at Eq.~(\ref{princ}) with $n=2$, where an extra exponent $e^{-i E_\alpha t}$ appears in the Schr\"{o}dinger representation.

Equation (\ref{ndip}) gives the transition amplitude to the leading order in $\omega_0^{-1}$ unless the term in the square brackets under the integral is independent on $\rv$. However, in the latter case the integration of $\hat{n}(\rv)$ over $\rv$ produces a constant $N$, and then the RHS becomes zero because of the zero the matrix element. This is, exactly, what happens if the field is uniform, as can be seen from Eqs.~(\ref{dipW}) and, therefore, this case requires a separate consideration.

\subsection{Uniform field case}

With the use of Eqs.~(\ref{AL}), (\ref{d3}), and with the commutator relations 
\begin{align}
&[\hat{H}_0, \sum\limits_{i=1}^N \mathbfcal{E}_0\cdot \rv_i]=
-\sum\limits_{i=1}^N \mathbfcal{E}_0 \cdot \nabla_i , \\
&[\hat{H}_0,[\hat{H}_0, \sum\limits_{i=1}^N \mathbfcal{E}_0\cdot \rv_i]] =
\sum\limits_{i=1}^N \mathbfcal{E}_0 \cdot \nabla_i v_{ext}(\rv_i), \\
\begin{split}
&[\hat{H}_0,[\hat{H}_0,[\hat{H}_0, \sum\limits_{i=1}^N \mathbfcal{E}_0 \cdot \rv_i]]] =
-\sum\limits_{i=1}^N \left\{ \frac{1}{2} \mathbfcal{E}_0 \cdot \nabla^3_i v_{ext}(\rv_i) \right. \\ &\left. + [\nabla_i (\mathbfcal{E}_0 \cdot \nabla_i v_{ext}(\rv_i) ]  \cdot \nabla_i\right\},
\end{split}
\\
&[[\hat{H}_0,[\hat{H}_0,[\hat{H}_0, \mathbfcal{E}_0 \cdot \rv]]],\sum\limits_{i=1}^N \mathbfcal{E}_0 \cdot\rv_i]= \\ 
&
- \sum\limits_{i=1}^N (\mathbfcal{E}_0 \cdot \nabla_i)^2 v_{ext}(\rv_i),  
\end{align}
and noting that in Eq.~(\ref{AL}) the sum of the 4th, 5th, and 6th terms on the RHS evaluates to zero, 
as it can be directly verified,
we immediately arrive at Eq.~(\ref{princ}) with $n=4$.

\subsection{Density oscillations and energy absorbed}

The time-dependent density is given by
\begin{widetext}
\begin{equation}
\begin{split}
&n(\rv,t>T)= \langle \Psi(t)|\hat{n}(\rv) | \Psi(t)\rangle 
=\sum\limits_{\alpha \beta} \langle \Psi_\alpha|\hat{n}(\rv)|\Psi_\beta\rangle
\langle \Psi(t)|\Psi_\alpha\rangle \langle \Psi_\beta|\Psi(t)\rangle = \langle \Psi_0|\hat{n}(\rv)|\Psi_0\rangle
|\langle \Psi_0|\Psi(t)\rangle|^2\\
&+
2 \, {\rm Re}\,\sum\limits_{\alpha\ne 0} \langle \Psi_0|\hat{n}(\rv)|\Psi_\alpha\rangle
\langle \Psi_\alpha|\Psi(t)\rangle \langle \Psi(t)|\Psi_0\rangle
 +
\sum\limits_{\alpha, \beta \ne 0 } \langle \Psi_\alpha|\hat{n}(\rv)|\Psi_\beta\rangle
\langle \Psi(t)|\Psi_\alpha\rangle \langle \Psi_\beta|\Psi(t)\rangle= |\langle \Psi_0|\hat{n}(\rv)|\Psi_0\rangle|^2\\
& -\langle \Psi_0|\hat{n}(\rv)|\Psi_0\rangle
\sum\limits_{\alpha\ne 0}|\langle \Psi_\alpha|\Psi(t)\rangle|^2+
2 \, {\rm Re}\,\sum\limits_{\alpha\ne 0} \langle \Psi_\alpha|\hat{n}(\rv)|\Psi_0\rangle
\langle \Psi_\alpha|\Psi(t)\rangle \langle \Psi(t)|\Psi_0\rangle
 +
\sum\limits_{\alpha, \beta \ne 0 } \langle \Psi_\alpha|\hat{n}(\rv)|\Psi_\beta\rangle
\langle \Psi(t)|\Psi_\alpha\rangle \langle \Psi_\beta|\Psi(t)\rangle,
\end{split}
\end{equation}
where the last equality is due to the normalization of $\Psi(t)$. Therefore,
\begin{equation}
\begin{split}
\delta n(\rv,t>T) &=  -\langle \Psi_0|\hat{n}(\rv)|\Psi_0\rangle
\! \sum\limits_{\alpha\ne 0}|\langle \Psi_\alpha|\Psi(t)\rangle|^2+
2 \, {\rm Re} \! \sum\limits_{\alpha\ne 0} \langle \Psi_0|\hat{n}(\rv)|\Psi_\alpha\rangle
\langle \Psi_\alpha|\Psi(t)\rangle \langle \Psi(t)|\Psi_0\rangle \\
  &+ 
\sum\limits_{\alpha, \beta \ne 0 } \langle \Psi_\alpha|\hat{n}(\rv)|\Psi_\beta\rangle
\langle \Psi(t)|\Psi_\alpha\rangle \langle \Psi_\beta|\Psi(t)\rangle.
\end{split}
\label{dnpr}
\end{equation}
\end{widetext}
With  account of Eq.~(\ref{princ}), we conclude that the leading term in $\omega_0^{-1}$ on RHS of Eq.~(\ref{dnpr}) is the second one,
while, for the same reason, $\langle \Psi(t)|\Psi_0\rangle=e^{i E_0 t}$ must be set in the latter. Then
\begin{equation}
\begin{split}
&\delta n(\rv,t>T)=  
2 \, {\rm Re} \, e^{i E_0 t} \sum\limits_{\alpha\ne 0} \langle \Psi_0|\hat{n}(\rv)|\Psi_\alpha\rangle
\langle \Psi_\alpha|\Psi(t)\rangle.
\end{split}
\label{dnpr2}
\end{equation}
Combining Eqs.~(\ref{princ}) and (\ref{dnpr2}), we have
\begin{equation}
\begin{split}
&\delta n(\rv,t>T)=  \frac{\pi }{ \omega_0^n}
 {\rm Re} \frac{1}{i} \! \sum\limits_{\alpha\ne 0} \langle \Psi_0|\hat{n}(\rv)|\Psi_\alpha\rangle
 \widetilde{C^2}(E_\alpha  -  E_0 ) \\
&\times e^{i (E_0-E_\alpha) t} \int \langle \Psi_\alpha | \hat{n}(\rv') |\Psi_0 \rangle  F_n(\rv')  d\rv',
\end{split}
\end{equation}
or
\begin{equation}
\begin{split}
&\delta n(\rv,t>T)=  \frac{\pi }{ \omega_0^n}
 {\rm Re} \frac{1}{i} \! \int e^{-i \omega t} \widetilde{C^2}(\omega ) \sum\limits_{\alpha\ne 0} \langle \Psi_0|\hat{n}(\rv)|\Psi_\alpha\rangle
 \\
&\times \langle \Psi_\alpha | \hat{n}(\rv') |\Psi_0 \rangle  F_n(\rv') \delta(\omega-E_\alpha+E_0) d\omega d\rv',
\end{split}
\label{dnpr3}
\end{equation}
Recalling the spectral representation of the many-body interacting density response function
\begin{equation}
\begin{split}
\chi(\rv,\rv',\omega) &=
\sum\limits_{\alpha\ne 0}
\left[
 \frac{\langle \Psi_\alpha|\hat{n}(\rv')|\Psi_0\rangle \langle \Psi_0|\hat{n}(\rv)|\Psi_\alpha\rangle}
{E_0-E_\alpha+\omega+i \eta}   \right. \\ &\left. +
\frac{\langle \Psi_\alpha|\hat{n}(\rv)|\Psi_0\rangle \langle \Psi_0|\hat{n}(\rv')|\Psi_\alpha\rangle}
{E_0-E_\alpha-\omega-i \eta}
\right],
\end{split}
\label{chi}
\end{equation}
where $\eta$ is a positive infinitesimal, 
we can rewrite Eq.~(\ref{dnpr3}) as
\begin{equation}
\begin{split}
&\delta n(\rv,t>T)=  \frac{\pi }{ \omega_0^n}
 {\rm Re} \frac{1}{i} \! \int e^{-i \omega t} \widetilde{C^2}(\omega ) 
 \\
&\times {\rm Im} \, \chi(\rv,\rv',\omega) F_n(\rv') d\omega d\rv'.
\end{split}
\end{equation}
Finally, the separation of the real part on the RHS of Eq.~(\ref{dnpr3}) can be dropped since the remaining expression is real already
(see the footnote \onlinecite{Note1}).

For the {\em total energy} absorbed by the system from the pulse, we can write
\begin{equation}
\Delta E= \sum\limits_{\alpha} E_\alpha  |\langle \Psi_\alpha|\Psi(t>T)\rangle|^2  -E_0,
\label{DE00}
\end{equation}
which, with the use of the completeness of the basis set,
can be rewritten as
\begin{equation}
\Delta E= \sum\limits_{\alpha\ne 0} (E_\alpha-E_0)  |\langle \Psi_\alpha|\Psi(t>T)\rangle|^2,
\label{DE0}
\end{equation}
and then, by Eq.~(\ref{princ}),
finally written in the form of Eq.~(\ref{DE}).

\

\section{Derivation in the Kramers-Henneberger's acceleration frame}
\label{AKH}
For an arbitrary $\uv(t)$, if a function $\Psi_{KH}(\{\rv\},t)$  satisfies the equation
\begin{equation}
\begin{split}
i \frac{\pa \Psi_{KH}(\{\rv\},t)}{\pa t}= 
\left\{ -\frac{1}{2}  \sum\limits_{i=1}^N   \nabla_i^2+ 
\frac{1}{2} \sum\limits_{i\ne j}^N \frac{1}{|\rv_i-\rv_j|}  \right. \\ \left. +
\sum\limits_{i=1}^N v_{ext}[\rv_i+\uv(t)] \right\} \Psi_{KH}(\{\rv\},t),
\end{split}
\label{KH}
\end{equation}
then the function
\begin{equation}
\Psi(\{\rv\},t)=e^{i\theta(\{\rv\},t)} \Psi_{KH}[\{\rv-\uv(t)\},t],
\label{SS4}
\end{equation}
where
\begin{equation}
\theta(\{\rv\},t) = \sum\limits_{i=1}^N \uv'(t)\cdot \rv_i ,
\end{equation}
satisfies the equation
\begin{equation}
\begin{split}
i \frac{\pa \Psi(\{\rv\},t)}{\pa t} = 
 \left\{ -\frac{1}{2}  \sum\limits_{i=1}^N   \nabla_i^2+ 
\frac{1}{2} \sum\limits_{i\ne j}^N \frac{1}{|\rv_i-\rv_j|}   \right. \\ \left.
+\sum\limits_{i=1}^N v_{ext}(\rv_i) -\sum\limits_{i=1}^N \uv''(t) \cdot \rv_i \right\} \Psi(\{\rv\},t). \\
\
 \end{split}
 \label{SS1}
\end{equation}
Choosing
\begin{equation}
\begin{split}
&\uv(t) =-\frac{\mathbfcal{E}_0 }{\omega_0^2} 
[ C(t) \cos\omega_0 t   + \\ & 
\int\limits_{-\infty}^t [(t-t') C''(t')-2 C'(t')] \cos \omega_0 t' d t' ]
\end{split}
\label{uv}
\end{equation}
and noting that
$\uv''(t)=\mathbfcal{E}_0 C(t) \cos \omega_0 t$,
we turn Eq.~(\ref{SS1}) into Eq.~(\ref{SE}) in the case of the dipole applied potential.

Expanding in Eq.~(\ref{KH}) up to $\omega_0^{-4}$, we have with the use of Eq.~(\ref{uv})
\begin{widetext}
\begin{equation}
\begin{split}
i \frac{\pa \Psi_{KH}(\{\rv\},t)}{\pa t}= \hat{H}_0 \Psi_{KH}(\{\rv\},t) 
&-\frac{C(t) \cos \omega_0 t }{\omega_0^2} 
\sum\limits_{i=1}^N [(\mathbfcal{E}_0\cdot\nabla_i) v_{ext}(\rv_i)]  \Psi_{KH}(\{\rv\},t) \\
&+\frac{C^2(t) \cos^2 \omega_0 t }{2 \omega_0^4} 
\sum\limits_{i=1}^N [(\mathbfcal{E}_0\cdot\nabla_i)^2 v_{ext}(\rv_i)]  \Psi_{KH}(\{\rv\},t),
\end{split}
\end{equation}
which in the interaction picture is written as
\begin{equation}
\begin{split}
i \frac{\pa \tilde{\Psi}_{KH}(\{\rv\},t)}{\pa t}= 
-\frac{C(t) \cos \omega_0 t }{\omega_0^2} 
\sum\limits_{i=1}^N e^{i \hat{H}_0 t}  [(\mathbfcal{E}_0\cdot\nabla_i) v_{ext}(\rv_i)]   e^{-i \hat{H}_0 t}  \tilde{\Psi}_{KH}(\{\rv\},t) \\
+\frac{C^2(t) \cos^2 \omega_0 t }{2 \omega_0^4} 
\sum\limits_{i=1}^N  e^{i \hat{H}_0 t}  [(\mathbfcal{E}_0\cdot\nabla_i)^2 v_{ext}(\rv_i)]   e^{-i \hat{H}_0 t}  \Psi_0(\{\rv\}),
\end{split}
\label{SS2}
\end{equation}
and, therefore,
\begin{equation}
\begin{split}
\tilde{\Psi}_{KH}(\{\rv\},+\infty)=\Psi_0(\{\rv\})-
\frac{1}{i \omega_0^2} \int\limits_{-\infty}^\infty e^{i \hat{H}_0 t'} \sum\limits_{i=1}^N [(\mathbfcal{E}_0\cdot\nabla_i) v_{ext}(\rv_i)] e^{-i \hat{H}_0 t'} \tilde{\Psi}_{KH}(\{\rv\},t') C(t') \cos \omega_0 t' d t'  \\
+\frac{1}{2 i \omega_0^4} \int\limits_{-\infty}^\infty e^{i \hat{H}_0 t'} \sum\limits_{i=1}^N 
 [(\mathbfcal{E}_0\cdot\nabla_i)^2 v_{ext}(\rv_i)] 
e^{-i \hat{H}_0 t'} \Psi_0(\{\rv\}) C^2(t') \cos^2 \omega_0 t' d t' .
\end{split}
\label{SS3}
\end{equation}
\end{widetext}
In the last terms on the RHS of Eqs.~(\ref{SS2}) and (\ref{SS3}) we have replaced $\tilde{\Psi}_{KH}(\{\rv\},t')$
with $\Psi_0(\{\rv\})$, which is in  accordance to keeping the terms up to $\omega_0^{-4}$ only.
We note that, upon the end of the pulse, according to Eqs.~(\ref{SS4}) and (\ref{uv}),
$\Psi_{KH}(\{\rv\},t)= \Psi(\{\rv\},t)$. Then,
the third term in the RHS of Eq.~(\ref{SS3}) immediately reproduces Eq.~(\ref{princ}). To prove that the contribution of the second term is zero up to $\omega_0^{-4}$ it is sufficient to integrate it by parts two times and use Eq.~(\ref{SS2}). 

\section{Particulars of the solution of the TD Schr\"{o}dinger equation for hydrogenic ion} 
\label{AConv}

In Eq.~(\ref{diff}), we expand $\tilde{\Psi}(\rv,t)$ as
\begin{equation}
\tilde{\Psi}(\rv,t) =\sum\limits_{l=0}^{l_{max}} \sum\limits_{n=0}^{n_{max}} a_{n,l}(t) F_n(r) Y_{l 0}(\theta,\phi),
\label{Aexp}
\end{equation}
where
\begin{align}
&F_n(r)= \lambda^{3/2} f_n(\lambda r),\\
&f_n(x)=\sqrt{\frac{n!}{\Gamma(n+\alpha+1)}}x^{\alpha/2-1}e^{-x/2} L^{(\alpha)}_n(x),
\end{align}
$L^{(\alpha)}_n(x)$ are the generalized Laguerre polynomials, and $\alpha$ and $\lambda$ are positive parameters. 
The basis set in Eq.~(\ref{Aexp}) is orthonormal and complete with any $\alpha$ and $\lambda$. Although we have been using $\alpha=2$ and $\lambda=1$, the convergence of the method has been verified by comparing results with those obtained with other values of these parameters. 

Matrix elements of the unperturbed Hamiltonian $\hat{H}_0$ and the time-dependent part $\hat{W}(t)$ were obtained exactly with the use of the recurrence relations for the generalized Laguerre polynomials. \cite{Abramowitz}
The problem was thus reduced to that of the propagation in time of the system of the linear ordinary differential equations for $a_{n,l}(t)$, which was carried out by means  of the Magnus expansion. \cite{Magnus-54}

For the hydrogen atom,
the Schr\"{o}dinger equation (\ref{SE}) reads
\begin{equation}
i \frac{\pa \Psi(\rv,t)}{\pa t}= \left[-\frac{1}{2} \nabla^2 -\frac{1}{r} + (\cos\omega_0 t) \hat{W}(\rv,t)\right] \Psi(\rv,t).
\label{SE1}
\end{equation}
By scaling the variables $\rv'= Z \rv$, $t'=Z^2 t$, we see that 
$\Psi_Z(\rv,t)= Z^{3/2} \Psi(Z \rv,Z^2 t)$ is the solution to the complementary problem for the hydrogenic atom of the nuclear charge $Z$
\begin{equation}
\begin{split}
i \frac{\pa \Psi_Z(\rv,t)}{\pa t} & = \left\{-\frac{1}{2} \nabla^2 -\frac{Z}{r} \right. \\
& \left. + Z^2[(\cos (Z^2 \omega_0 t)] \hat{W}(Z \rv,Z^2 t)\right\} \Psi_Z(\rv,t).
\end{split}
\label{SEZ}
\end{equation}
From Eq.~(\ref{SEZ}) we conclude that the frequency $\omega_0$ scales as $\omega_0 \to Z^2 \omega_0$,
which explains the faster convergence of the solutions to its $\omega_0 \to \infty$ limit we have observed  in Fig.~\ref{Z025w8}
for $Z<1$. 

\section{TDDFT calculation of ethylene spectrum}
\label{AR}
The energy absorption spectra were calculated using time-dependent local density approximation performed in real-time on a real-space grid. We used Troullier-Martins norm-conserving pseudopotentials \cite{troullier_efficient_1991} and the reciprocal-space-based method for treating long-range interactions. \cite{martyna_reciprocal_1999} The molecule C-C axis coincides with the z-axis, and the four hydrogen atoms are in the y-z place. A local density approximation energy minimization determined the atom distance. The time propagation used fourth-order Runge-Kutta propagation with a time step of 0.05 atomic time units. 

%\bibliography{ref}

%aipnum4-2.bst 2019-01-14 (MD) hand-edited version of apsrev4-1.bst

\onecolumngrid
\newpage
\thispagestyle{empty}

\renewcommand{\theequation}{{S.\arabic{equation}}}
\renewcommand{\thefigure}{{S.\arabic{figure}}}

\section*{Supplementary Material \\ to the article 
'High-frequency limit of spectroscopy' by Vladimir~U.~Nazarov and Roi Baer}
\subsection*{Derivation of Eq.~(\ref{AL}).}

From Eq.~(\ref{inteq}), by the integration by parts, we can write
\begin{equation}
\tilde{ \Psi}(t) = \Psi_0 + \frac{1}{i\omega_0} \int\limits_{-\infty}^t 
 \tilde{\hat{W}}(t')
 \tilde{\Psi}(t')  d \sin \omega_0 t'= \Psi_0 + \frac{1}{i\omega_0}  
 \tilde{\hat{W}}(t)
 \tilde{\Psi}(t)  \sin \omega_0 t-\frac{1}{i\omega_0} \int\limits_{-\infty}^t 
 \frac{\pa}{\pa t'} \left[\tilde{\hat{W}}(t')
 \tilde{\Psi}(t')\right]  \sin \omega_0 t' d t',
 \label{pre}
\end{equation}
or
\begin{equation}
\tilde{ \Psi}(t) = \Psi_0 + \frac{1}{i\omega_0}  
 \tilde{\hat{W}}(t)
 \tilde{\Psi}(t)  \sin \omega_0 t-\frac{1}{i\omega_0} \int\limits_{-\infty}^t 
 \left[ \frac{\pa}{\pa t'} \tilde{\hat{W}}(t') \right]
 \tilde{\Psi}(t')  \sin \omega_0 t' d t'
 -\frac{1}{i\omega_0} \int\limits_{-\infty}^t 
\tilde{\hat{W}}(t') 
 \left[ \frac{\pa}{\pa t'} \tilde{\Psi}(t')\right]  \sin \omega_0 t' d t',
\end{equation}
and, with the use of Eq.~(\ref{diff}),
\begin{equation}
\tilde{ \Psi}(t) = \Psi_0 + \frac{1}{i\omega_0}  
 \tilde{\hat{W}}(t)
 \tilde{\Psi}(t)  \sin \omega_0 t-\frac{1}{i\omega_0} \int\limits_{-\infty}^t 
 \left[ \frac{\pa}{\pa t'} \tilde{\hat{W}}(t') \right]
 \tilde{\Psi}(t')  \sin \omega_0 t' d t'
 +\frac{1}{2 \omega_0} \int\limits_{-\infty}^t 
\tilde{\hat{W}}^2(t') 
 \tilde{\Psi}(t') \sin 2 \omega_0 t' d t'.
\end{equation}
Continuing in the same way
\begin{equation}
\tilde{ \Psi}(t) = \Psi_0 + \frac{1}{i\omega_0}  
 \tilde{\hat{W}}(t)
 \tilde{\Psi}(t)  \sin \omega_0 t+\frac{1}{i\omega_0^2} \int\limits_{-\infty}^t 
 \left[ \frac{\pa}{\pa t'} \tilde{\hat{W}}(t') \right]
 \tilde{\Psi}(t')  d \cos \omega_0 t' 
 -\frac{1}{4 \omega_0^2} \int\limits_{-\infty}^t 
\tilde{\hat{W}}^2(t') 
 \tilde{\Psi}(t') d \cos 2 \omega_0 t' ,
\end{equation}

\begin{equation}
\begin{split}
\tilde{ \Psi}(t) = \Psi_0 + \frac{1}{i\omega_0}  
 \tilde{\hat{W}}(t)
 \tilde{\Psi}(t)  \sin \omega_0 t
+\frac{1}{i\omega_0^2} 
 \left[ \frac{\pa}{\pa t} \tilde{\hat{W}}(t) \right]
 \tilde{\Psi}(t)  \cos \omega_0 t
 -\frac{1}{4 \omega_0^2} 
\tilde{\hat{W}}^2(t) 
 \tilde{\Psi}(t)  \cos 2 \omega_0 t \\
-\frac{1}{i\omega_0^2} \int\limits_{-\infty}^t 
\frac{\pa}{\pa t'} \left\{\left[ \frac{\pa}{\pa t'} \tilde{\hat{W}}(t') \right]
 \tilde{\Psi}(t') \right\}  \cos \omega_0 t' d t'
 +\frac{1}{4 \omega_0^2} \int\limits_{-\infty}^t 
\frac{\pa}{\pa t'} \left\{\tilde{\hat{W}}^2(t') 
 \tilde{\Psi}(t)\right\}   \cos 2 \omega_0 t' d t',
 \end{split}
\end{equation}

\begin{equation}
\begin{split}
\tilde{ \Psi}(t) = \Psi_0 + \frac{1}{i\omega_0}  
 \tilde{\hat{W}}(t)
 \tilde{\Psi}(t)  \sin \omega_0 t
+\frac{1}{i\omega_0^2} 
 \left[ \frac{\pa}{\pa t} \tilde{\hat{W}}(t) \right]
 \tilde{\Psi}(t)  \cos \omega_0 t
 -\frac{1}{4 \omega_0^2} 
\tilde{\hat{W}}^2(t) 
 \tilde{\Psi}(t)  \cos 2 \omega_0 t \\
-\frac{1}{i\omega_0^2} \int\limits_{-\infty}^t 
 \left\{\left[ \frac{\pa^2}{\pa t'^2} \tilde{\hat{W}}(t') \right]
 \tilde{\Psi}(t') \right\}  \cos \omega_0 t' d t'
 -\frac{1}{i\omega_0^2} \int\limits_{-\infty}^t 
 \left[ \frac{\pa}{\pa t'} \tilde{\hat{W}}(t') \right]
 \left[\frac{\pa}{\pa t'}
 \tilde{\Psi}(t') \right]  \cos \omega_0 t' d t' \\
 +\frac{1}{4 \omega_0^2} \int\limits_{-\infty}^t 
 \left[\frac{\pa}{\pa t'} \tilde{\hat{W}}^2(t') \right]
 \tilde{\Psi}(t)   \cos 2 \omega_0 t' d t'
 +\frac{1}{4 \omega_0^2} \int\limits_{-\infty}^t 
 \tilde{\hat{W}}^2(t') 
\left[\frac{\pa}{\pa t'}  \tilde{\Psi}(t) \right]  \cos 2 \omega_0 t' d t',
 \end{split}
\end{equation}

\begin{equation}
\begin{split}
&\tilde{ \Psi}(t) = \Psi_0 + \frac{1}{i\omega_0}  
 \tilde{\hat{W}}(t)
 \tilde{\Psi}(t)  \sin \omega_0 t
+\frac{1}{i\omega_0^2} 
 \left[ \frac{\pa}{\pa t} \tilde{\hat{W}}(t) \right]
 \tilde{\Psi}(t)  \cos \omega_0 t
 -\frac{1}{4 \omega_0^2} 
\tilde{\hat{W}}^2(t) 
 \tilde{\Psi}(t)  \cos 2 \omega_0 t \\
&-\frac{1}{i\omega_0^2} \int\limits_{-\infty}^t 
 \left[ \frac{\pa^2}{\pa t'^2} \tilde{\hat{W}}(t') \right]
 \tilde{\Psi}(t')  \cos \omega_0 t' d t'
 +\frac{1}{\omega_0^2} \int\limits_{-\infty}^t 
 \left[ \frac{\pa}{\pa t'} \tilde{\hat{W}}(t') \right]
 \tilde{\hat{W}}(t') \tilde{\Psi}(t') \cos^2 \omega_0 t' d t' \\
& +\frac{1}{4 \omega_0^2} \int\limits_{-\infty}^t 
 \left[\frac{\pa}{\pa t'} \tilde{\hat{W}}^2(t') \right]
 \tilde{\Psi}(t')   \cos 2 \omega_0 t' d t'
 +\frac{1}{4 i \omega_0^2} \int\limits_{-\infty}^t 
 \tilde{\hat{W}}^3(t') 
\tilde{\Psi}(t')  \cos\omega_0 t \cos 2 \omega_0 t' d t'.
\label{imp}
 \end{split}
\end{equation}

Since
\begin{equation}
\left[\frac{\pa}{\pa t}\tilde{\hat{W}}(t) \right] \tilde{\hat{W}}(t)=
\frac{1}{2}\frac{\pa}{\pa t}\tilde{\hat{W}}^2(t) + \frac{1}{2}\left[\frac{\pa}{\pa t}\tilde{\hat{W}}(t), \tilde{\hat{W}}(t) \right],
\end{equation}
we can rewrite Eq.~(\ref{imp}) as
\begin{equation}
\begin{split}
&\tilde{ \Psi}(t) \! = \! \Psi_0 \! + \! \frac{1}{i\omega_0}  
 \tilde{\hat{W}}(t)
 \tilde{\Psi}(t)  \sin \omega_0 t
\! + \! \frac{1}{i\omega_0^2} 
 \left[ \frac{\pa}{\pa t} \tilde{\hat{W}}(t) \right]
 \tilde{\Psi}(t)  \cos \omega_0 t
 \! - \! \frac{1}{4 \omega_0^2} 
\tilde{\hat{W}}^2(t) 
 \tilde{\Psi}(t)  \cos 2 \omega_0 t 
\! - \! \frac{1}{i\omega_0^2} \! \int\limits_{-\infty}^t \!
 \left[ \frac{\pa^2}{\pa t'^2} \tilde{\hat{W}}(t') \right]
 \! \tilde{\Psi}(t')  \cos \omega_0 t' d t' \\
 &+\frac{1}{4\omega_0^2} \int\limits_{-\infty}^t 
 \left[ \frac{\pa}{\pa t'} \tilde{\hat{W}}^2(t') \right]
 \tilde{\Psi}(t')  d t' +\frac{1}{4\omega_0^2} \int\limits_{-\infty}^t 
 \left[ \frac{\pa}{\pa t'} \tilde{\hat{W}}(t') ,
 \tilde{\hat{W}}(t') \right]\tilde{\Psi}(t')  d t'+\frac{1}{4 \omega_0^2} \int\limits_{-\infty}^t 
 \left[ \frac{\pa}{\pa t'} \tilde{\hat{W}}(t') ,
 \tilde{\hat{W}}(t')\right] \tilde{\Psi}(t') \cos 2 \omega_0 t' d t'\\
& +\frac{1}{2 \omega_0^2} \int\limits_{-\infty}^t 
 \left[\frac{\pa}{\pa t'} \tilde{\hat{W}}^2(t') \right]
 \tilde{\Psi}(t')   \cos 2 \omega_0 t' d t'
 +\frac{1}{4 i \omega_0^2} \int\limits_{-\infty}^t 
 \tilde{\hat{W}}^3(t') 
\tilde{\Psi}(t')  \cos\omega_0 t \cos 2 \omega_0 t' d t'.
 \end{split}
\end{equation}
Furthermore
\begin{equation}
\begin{split}
&\tilde{ \Psi}(t) = \Psi_0 + \frac{1}{i\omega_0}  
 \tilde{\hat{W}}(t)
 \tilde{\Psi}(t)  \sin \omega_0 t
+\frac{1}{i\omega_0^2} 
 \left[ \frac{\pa}{\pa t} \tilde{\hat{W}}(t) \right]
 \tilde{\Psi}(t)  \cos \omega_0 t
 -\frac{1}{4 \omega_0^2} 
\tilde{\hat{W}}^2(t) 
 \tilde{\Psi}(t)  \cos 2 \omega_0 t +\frac{1}{4\omega_0^2} 
   \tilde{\hat{W}}^2(t) \tilde{\Psi}(t)  \\
&-\frac{1}{i\omega_0^2} \int\limits_{-\infty}^t 
 \left[ \frac{\pa^2}{\pa t'^2} \tilde{\hat{W}}(t') \right]
 \tilde{\Psi}(t')  \cos \omega_0 t' d t' -\frac{1}{4\omega_0^2} \int\limits_{-\infty}^t 
 \tilde{\hat{W}}^2(t') 
 \left[ \frac{\pa}{\pa t'} \tilde{\Psi}(t') \right] d t'\\
 & +\frac{1}{4\omega_0^2} \int\limits_{-\infty}^t 
 \left[ \frac{\pa}{\pa t'} \tilde{\hat{W}}(t') ,
 \tilde{\hat{W}}(t') \right]\tilde{\Psi}(t')  d t'+\frac{1}{4 \omega_0^2} \int\limits_{-\infty}^t 
 \left[ \frac{\pa}{\pa t'} \tilde{\hat{W}}(t') ,
 \tilde{\hat{W}}(t')\right] \tilde{\Psi}(t') \cos 2 \omega_0 t' d t'\\
& +\frac{1}{2 \omega_0^2} \int\limits_{-\infty}^t 
 \left[\frac{\pa}{\pa t'} \tilde{\hat{W}}^2(t') \right]
 \tilde{\Psi}(t')   \cos 2 \omega_0 t' d t'
 +\frac{1}{8 i \omega_0^2} \int\limits_{-\infty}^t 
 \tilde{\hat{W}}^3(t') 
\tilde{\Psi}(t')  (\cos\omega_0 t +\cos 3 \omega_0 t') d t',
 \end{split}
\end{equation}
\begin{equation}
\begin{split}
&\tilde{ \Psi}(t) = \Psi_0 + \frac{1}{i\omega_0}  
 \tilde{\hat{W}}(t)
 \tilde{\Psi}(t)  \sin \omega_0 t
+\frac{1}{i\omega_0^2} 
 \left[ \frac{\pa}{\pa t} \tilde{\hat{W}}(t) \right]
 \tilde{\Psi}(t)  \cos \omega_0 t
 -\frac{1}{4 \omega_0^2} 
\tilde{\hat{W}}^2(t) 
 \tilde{\Psi}(t)  \cos 2 \omega_0 t +\frac{1}{4\omega_0^2} 
   \tilde{\hat{W}}^2(t) \tilde{\Psi}(t) \\
&-\frac{1}{i\omega_0^2} \! \int\limits_{-\infty}^t \!
 \left[ \frac{\pa^2}{\pa t'^2} \tilde{\hat{W}}(t') \right]
\! \tilde{\Psi}(t')  \cos \omega_0 t' d t' 
\! + \! \frac{1}{4\omega_0^2} \! \int\limits_{-\infty}^t \!
 \left[ \frac{\pa}{\pa t'} \tilde{\hat{W}}(t') ,
 \tilde{\hat{W}}(t') \right] \! \tilde{\Psi}(t')  d t' \! + \! \frac{1}{4 \omega_0^2} \! \int\limits_{-\infty}^t \!
 \left[ \frac{\pa}{\pa t'} \tilde{\hat{W}}(t') ,
 \tilde{\hat{W}}(t')\right] \! \tilde{\Psi}(t') \cos 2 \omega_0 t' d t'\\
& +\frac{1}{2 \omega_0^2} \int\limits_{-\infty}^t 
 \left[\frac{\pa}{\pa t'} \tilde{\hat{W}}^2(t') \right]
 \tilde{\Psi}(t')   \cos 2 \omega_0 t' d t'
 +\frac{1}{8 i \omega_0^2} \int\limits_{-\infty}^t 
 \tilde{\hat{W}}^3(t') 
\tilde{\Psi}(t')  ( \cos 3 \omega_0 t'-\cos\omega_0 t') d t',
 \end{split}
\end{equation}
\begin{equation}
\begin{split}
&\tilde{ \Psi}(t) = \Psi_0 + \frac{1}{i\omega_0}  
 \tilde{\hat{W}}(t)
 \tilde{\Psi}(t)  \sin \omega_0 t
+\frac{1}{i\omega_0^2} 
 \left[ \frac{\pa}{\pa t} \tilde{\hat{W}}(t) \right]
 \tilde{\Psi}(t)  \cos \omega_0 t
 -\frac{1}{4 \omega_0^2} 
\tilde{\hat{W}}^2(t) 
 \tilde{\Psi}(t)  \cos 2 \omega_0 t +\frac{1}{4\omega_0^2} 
   \tilde{\hat{W}}^2(t) \tilde{\Psi}(t) \\
&-\frac{1}{i\omega_0^3} \int\limits_{-\infty}^t 
 \left[ \frac{\pa^2}{\pa t'^2} \tilde{\hat{W}}(t') \right]
 \tilde{\Psi}(t')  d \sin \omega_0 t'  
  +\frac{1}{4\omega_0^2} \int\limits_{-\infty}^t 
 \left[ \frac{\pa}{\pa t'} \tilde{\hat{W}}(t') ,
 \tilde{\hat{W}}(t') \right]\tilde{\Psi}(t')  d t' +\frac{1}{8 \omega_0^3} \int\limits_{-\infty}^t 
 \left[ \frac{\pa}{\pa t'} \tilde{\hat{W}}(t') ,
 \tilde{\hat{W}}(t')\right] \tilde{\Psi}(t') d \sin 2 \omega_0 t' \\ 
 &
 +\frac{1}{4 \omega_0^3} \int\limits_{-\infty}^t 
 \left[\frac{\pa}{\pa t'} \tilde{\hat{W}}^2(t') \right]
 \tilde{\Psi}(t')   d\sin 2 \omega_0 t' 
 +\frac{1}{8 i \omega_0^3} \int\limits_{-\infty}^t 
 \tilde{\hat{W}}^3(t') 
\tilde{\Psi}(t')  ( \frac{1}{3} d\sin 3 \omega_0 t'-d\sin \omega_0 t'),
 \end{split}
\end{equation} 
\begin{equation}
\begin{split}
&\tilde{ \Psi}(t) = \Psi_0 + \frac{1}{i\omega_0}  
 \tilde{\hat{W}}(t)
 \tilde{\Psi}(t)  \sin \omega_0 t
+\frac{1}{i\omega_0^2} 
 \left[ \frac{\pa}{\pa t} \tilde{\hat{W}}(t) \right]
 \tilde{\Psi}(t)  \cos \omega_0 t
 -\frac{1}{4 \omega_0^2} 
\tilde{\hat{W}}^2(t) 
 \tilde{\Psi}(t)  \cos 2 \omega_0 t +\frac{1}{4\omega_0^2} 
   \tilde{\hat{W}}^2(t) \tilde{\Psi}(t) \\
&-\frac{1}{i\omega_0^3}
 \left[ \frac{\pa^2}{\pa t^2} \tilde{\hat{W}}(t) \right]
 \tilde{\Psi}(t)   \sin \omega_0 t  
  +\frac{1}{8 \omega_0^3} 
 \left[ \frac{\pa}{\pa t} \tilde{\hat{W}}(t) ,
 \tilde{\hat{W}}(t)\right] \tilde{\Psi}(t)  \sin 2 \omega_0 t \\
& +\frac{1}{4 \omega_0^3} 
 \left[\frac{\pa}{\pa t} \tilde{\hat{W}}^2(t) \right]
 \tilde{\Psi}(t)   \sin 2 \omega_0 t 
 +\frac{1}{8 i \omega_0^3} 
 \tilde{\hat{W}}^3(t) 
\tilde{\Psi}(t)  ( \frac{1}{3} \sin 3 \omega_0 t-\sin \omega_0 t)
+\frac{1}{i\omega_0^3} \int\limits_{-\infty}^t 
\frac{\pa}{\pa t'} \left\{ \left[ \frac{\pa^2}{\pa t'^2} \tilde{\hat{W}}(t') \right]
 \tilde{\Psi}(t') \right\}  \sin \omega_0 t' d t' \\
 & +\frac{1}{4\omega_0^2} \int\limits_{-\infty}^t 
 \left[ \frac{\pa}{\pa t'} \tilde{\hat{W}}(t') ,
 \tilde{\hat{W}}(t') \right]\tilde{\Psi}(t')  d t'-\frac{1}{8 \omega_0^3} \int\limits_{-\infty}^t 
\left\{  \frac{\pa}{\pa t'} \left[ \frac{\pa}{\pa t'} \tilde{\hat{W}}(t') ,
 \tilde{\hat{W}}(t')\right] \tilde{\Psi}(t') \right\}  \sin 2 \omega_0 t' d t' \\
& -\frac{1}{4 \omega_0^3} \int\limits_{-\infty}^t 
\left\{ \frac{\pa}{\pa t'} \left[\frac{\pa}{\pa t'} \tilde{\hat{W}}^2(t') \right]
 \tilde{\Psi}(t') \right\}  \sin 2 \omega_0 t' d t'
 -\frac{1}{8 i \omega_0^3} \int\limits_{-\infty}^t 
\left\{\frac{\pa}{\pa t'} \tilde{\hat{W}}^3(t') 
\tilde{\Psi}(t') \right\} ( \frac{1}{3} \sin 3 \omega_0 t'-\sin \omega_0 t') d t',
 \end{split}
\end{equation}
\begin{equation}
\begin{split}
&\tilde{ \Psi}(t) = \Psi_0 + \frac{1}{i\omega_0}  
 \tilde{\hat{W}}(t)
 \tilde{\Psi}(t)  \sin \omega_0 t
+\frac{1}{i\omega_0^2} 
 \left[ \frac{\pa}{\pa t} \tilde{\hat{W}}(t) \right]
 \tilde{\Psi}(t)  \cos \omega_0 t
 -\frac{1}{4 \omega_0^2} 
\tilde{\hat{W}}^2(t) 
 \tilde{\Psi}(t)  \cos 2 \omega_0 t +\frac{1}{4\omega_0^2} 
   \tilde{\hat{W}}^2(t) \tilde{\Psi}(t) \\
&-\frac{1}{i\omega_0^3}
 \left[ \frac{\pa^2}{\pa t^2} \tilde{\hat{W}}(t) \right]
 \tilde{\Psi}(t)   \sin \omega_0 t  
  +\frac{1}{8 \omega_0^3} 
 \left[ \frac{\pa}{\pa t} \tilde{\hat{W}}(t) ,
 \tilde{\hat{W}}(t)\right] \tilde{\Psi}(t)  \sin 2 \omega_0 t \\
& +\frac{1}{4 \omega_0^3} 
 \left[\frac{\pa}{\pa t} \tilde{\hat{W}}^2(t) \right]
 \tilde{\Psi}(t)   \sin 2 \omega_0 t 
 +\frac{1}{8 i \omega_0^3} 
 \tilde{\hat{W}}^3(t) 
\tilde{\Psi}(t)  ( \frac{1}{3} \sin 3 \omega_0 t-\sin \omega_0 t)\\
&+\frac{1}{i\omega_0^3} \int\limits_{-\infty}^t 
\left[ \frac{\pa^3}{\pa t'^3} \tilde{\hat{W}}(t') \right]
 \tilde{\Psi}(t')   \sin \omega_0 t' d t' -\frac{1}{2\omega_0^3} \int\limits_{-\infty}^t 
 \left[ \frac{\pa^2}{\pa t'^2} \tilde{\hat{W}}(t') \right]
 \tilde{\hat{W}}(t')
 \tilde{\Psi}(t')  \sin 2\omega_0 t' d t'
 +\frac{1}{4\omega_0^2} \int\limits_{-\infty}^t 
 \left[ \frac{\pa}{\pa t'} \tilde{\hat{W}}(t') ,
 \tilde{\hat{W}}(t') \right]\tilde{\Psi}(t')  d t'  \\ 
 &-\frac{1}{8 \omega_0^3} \int\limits_{-\infty}^t 
 \left[ \frac{\pa^2}{\pa t'^2} \tilde{\hat{W}}(t') ,
 \tilde{\hat{W}}(t')\right] \tilde{\Psi}(t')  \sin 2 \omega_0 t' d t' -\frac{1}{16 i\omega_0^3} \int\limits_{-\infty}^t 
  \left[ \frac{\pa}{\pa t'} \tilde{\hat{W}}(t') ,
 \tilde{\hat{W}}(t')\right] 
\tilde{\hat{W}}(t')
 \tilde{\Psi}(t')  (\sin \omega_0 t'+\sin 3\omega_0 t')d t' \\
& -\frac{1}{4 \omega_0^3} \int\limits_{-\infty}^t 
  \left[\frac{\pa^2}{\pa t'^2} \tilde{\hat{W}}^2(t') \right]
 \tilde{\Psi}(t')   \sin 2 \omega_0 t' d t'-\frac{1}{8 i\omega_0^3} \int\limits_{-\infty}^t 
  \left[\frac{\pa}{\pa t'} \tilde{\hat{W}}^2(t') \right]
 \tilde{\hat{W}}(t')
 \tilde{\Psi}(t') (\sin \omega_0 t'+\sin 3\omega_0 t') d t' \\
& -\frac{1}{8 i \omega_0^3} \int\limits_{-\infty}^t 
\left[\frac{\pa}{\pa t'} \tilde{\hat{W}}^3(t') \right]
\tilde{\Psi}(t')  ( \frac{1}{3} \sin 3 \omega_0 t'-\sin \omega_0 t') d t'+\frac{1}{48 \omega_0^3} \int\limits_{-\infty}^t 
 \tilde{\hat{W}}^4(t') 
 \tilde{\Psi}(t')  (\sin 4 \omega_0 t'-2 \sin 2 \omega_0 t')d t',
 \end{split}
\end{equation}
\allowdisplaybreaks
%\begin{equation}
%\begin{split}
\begin{align}
&\tilde{ \Psi}(t) = \Psi_0 + \frac{1}{i\omega_0}  
 \tilde{\hat{W}}(t)
 \tilde{\Psi}(t)  \sin \omega_0 t
+\frac{1}{i\omega_0^2} 
 \left[ \frac{\pa}{\pa t} \tilde{\hat{W}}(t) \right]
 \tilde{\Psi}(t)  \cos \omega_0 t
 -\frac{1}{4 \omega_0^2} 
\tilde{\hat{W}}^2(t) 
 \tilde{\Psi}(t)  \cos 2 \omega_0 t +\frac{1}{4\omega_0^2} 
   \tilde{\hat{W}}^2(t) \tilde{\Psi}(t) \nonumber \\
&-\frac{1}{i\omega_0^3}
 \left[ \frac{\pa^2}{\pa t^2} \tilde{\hat{W}}(t) \right]
 \tilde{\Psi}(t)   \sin \omega_0 t  
  +\frac{1}{8 \omega_0^3} 
 \left[ \frac{\pa}{\pa t} \tilde{\hat{W}}(t) ,
 \tilde{\hat{W}}(t)\right] \tilde{\Psi}(t)  \sin 2 \omega_0 t \nonumber \\
& +\frac{1}{4 \omega_0^3} 
 \left[\frac{\pa}{\pa t} \tilde{\hat{W}}^2(t) \right]
 \tilde{\Psi}(t)   \sin 2 \omega_0 t 
 +\frac{1}{8 i \omega_0^3} 
 \tilde{\hat{W}}^3(t) 
\tilde{\Psi}(t)  ( \frac{1}{3} \sin 3 \omega_0 t-\sin \omega_0 t) \nonumber \\
& -\frac{1}{i\omega_0^4} 
\left[ \frac{\pa^3}{\pa t^3} \tilde{\hat{W}}(t) \right]
 \tilde{\Psi}(t)    \cos \omega_0 t +\frac{1}{4\omega_0^4} 
 \left[ \frac{\pa^2}{\pa t^2} \tilde{\hat{W}}(t) \right]
 \tilde{\hat{W}}(t)
 \tilde{\Psi}(t)   \cos 2\omega_0 t \nonumber \\ 
 &+\frac{1}{16 \omega_0^4} 
 \left[ \frac{\pa^2}{\pa t^2} \tilde{\hat{W}}(t) ,
 \tilde{\hat{W}}(t)\right] \tilde{\Psi}(t)   \cos 2 \omega_0 t  +\frac{1}{16 i\omega_0^4} 
  \left[ \frac{\pa}{\pa t} \tilde{\hat{W}}(t) ,
 \tilde{\hat{W}}(t)\right] 
\tilde{\hat{W}}(t)
 \tilde{\Psi}(t)  ( \cos \omega_0 t+\frac{1}{3} \cos 3\omega_0 t) \nonumber \\
& +\frac{1}{8 \omega_0^4} 
  \left[\frac{\pa^2}{\pa t^2} \tilde{\hat{W}}^2(t) \right]
 \tilde{\Psi}(t)    \cos 2 \omega_0 t +\frac{1}{8 i\omega_0^4} 
  \left[\frac{\pa}{\pa t} \tilde{\hat{W}}^2(t) \right]
 \tilde{\hat{W}}(t)
 \tilde{\Psi}(t) ( \cos \omega_0 t+\frac{1}{3}  \cos 3\omega_0 t)  \nonumber \\
& +\frac{1}{8 i \omega_0^4} 
\left[\frac{\pa}{\pa t} \tilde{\hat{W}}^3(t) \right]
\tilde{\Psi}(t)  ( \frac{1}{9}  \cos 3 \omega_0 t- \cos \omega_0 t) -\frac{1}{48 \omega_0^4} 
 \tilde{\hat{W}}^4(t) 
 \tilde{\Psi}(t)  ( \frac{1}{4}  \cos 4 \omega_0 t- \cos 2 \omega_0 t) \nonumber \\ 
 &+\frac{1}{4\omega_0^2} \int\limits_{-\infty}^t 
 \left[ \frac{\pa}{\pa t'} \tilde{\hat{W}}(t') ,
 \tilde{\hat{W}}(t') \right]\tilde{\Psi}(t')  d t' +\frac{1}{i\omega_0^4} \int\limits_{-\infty}^t 
 \left[ \frac{\pa^4}{\pa t'^4} \tilde{\hat{W}}(t') \right]
 \tilde{\Psi}(t')    \cos \omega_0 t' d t' -\frac{1}{\omega_0^4} \int\limits_{-\infty}^t 
 \left[ \frac{\pa^3}{\pa t'^3} \tilde{\hat{W}}(t') \right]
 \tilde{\hat{W}}(t) \tilde{\Psi}(t)   \cos^2 \omega_0 t' d t' \nonumber \\
 &-\frac{1}{4\omega_0^4} \int\limits_{-\infty}^t 
\left\{\frac{\pa}{\pa t'} \left\{ \left[ \frac{\pa^2}{\pa t'^2} \tilde{\hat{W}}(t') \right]
 \tilde{\hat{W}}(t')\right\}\right\}
 \tilde{\Psi}(t')   \cos 2\omega_0 t' d t'-\frac{1}{4i \omega_0^4} \int\limits_{-\infty}^t 
 \left[ \frac{\pa^2}{\pa t'^2} \tilde{\hat{W}}(t') \right]
 \tilde{\hat{W}}^2(t')
 \tilde{\Psi}(t') \cos\omega_0 t \cos 2\omega_0 t' d t' \nonumber \\ 
 &-\frac{1}{16 \omega_0^4} \int\limits_{-\infty}^t 
\left\{ \frac{\pa}{\pa t'}  \left[ \frac{\pa^2}{\pa t'^2} \tilde{\hat{W}}(t') ,
 \tilde{\hat{W}}(t')\right]\right\} \tilde{\Psi}(t')   \cos 2 \omega_0 t' d t'-\frac{1}{16 i \omega_0^4} \int\limits_{-\infty}^t 
  \left[ \frac{\pa^2}{\pa t'^2} \tilde{\hat{W}}(t') ,
 \tilde{\hat{W}}(t')\right]   \tilde{\hat{W}}(t') \tilde{\Psi}(t')  \cos\omega_0 t' \cos 2 \omega_0 t' d t' \nonumber \\
 &-\frac{1}{16 i\omega_0^4} \int\limits_{-\infty}^t 
\left\{\frac{\pa}{\pa t'}    \left\{\left[ \frac{\pa}{\pa t'} \tilde{\hat{W}}(t') ,
 \tilde{\hat{W}}(t')\right] 
\tilde{\hat{W}}(t')\right\}\right\}
 \tilde{\Psi}(t')  ( \cos \omega_0 t'+\frac{1}{3} \cos 3\omega_0 t')d t' \nonumber \\
 &+\frac{1}{16 \omega_0^4} \int\limits_{-\infty}^t 
   \left[ \frac{\pa}{\pa t'} \tilde{\hat{W}}(t') ,
 \tilde{\hat{W}}(t')\right] 
\tilde{\hat{W}}^2(t')
 \tilde{\Psi}(t)  ( \cos^2 \omega_0 t'+\frac{1}{3} \cos\omega_0 t'\cos 3\omega_0 t')d t' \nonumber \\
& -\frac{1}{8 \omega_0^4} \int\limits_{-\infty}^t 
  \left[\frac{\pa^3}{\pa t'^3} \tilde{\hat{W}}^2(t') \right]
 \tilde{\Psi}(t')    \cos 2 \omega_0 t' d t' -\frac{1}{8 i \omega_0^4} \int\limits_{-\infty}^t 
  \left[\frac{\pa^2}{\pa t'^2} \tilde{\hat{W}}^2(t') \right]
  \tilde{\hat{W}}(t') \tilde{\Psi}(t')   \cos\omega_0 t' \cos 2 \omega_0 t' d t' \nonumber \\
 &-\frac{1}{8 i\omega_0^4} \int\limits_{-\infty}^t 
\left\{\frac{\pa}{\pa t'} \left\{ \left[\frac{\pa}{\pa t'} \tilde{\hat{W}}^2(t') \right]
 \tilde{\hat{W}}(t')\right\}\right\}
 \tilde{\Psi}(t') ( \cos \omega_0 t'+\frac{1}{3}  \cos 3\omega_0 t') d t' \nonumber \\
 &+\frac{1}{8 \omega_0^4} \int\limits_{-\infty}^t 
  \left[\frac{\pa}{\pa t'} \tilde{\hat{W}}^2(t') \right]
 \tilde{\hat{W}}^2(t')
  \tilde{\Psi}(t')   ( \cos^2 \omega_0 t'+\frac{1}{3} \cos\omega_0 t' \cos 3\omega_0 t') d t' 
  %\tag{\stepcounter{equation}\theequation}  
  \label{ttag} \\
& -\frac{1}{8 i \omega_0^4} \int\limits_{-\infty}^t 
 \left[\frac{\pa^2}{\pa t'^2} \tilde{\hat{W}}^3(t') \right]
\tilde{\Psi}(t')  ( \frac{1}{9}  \cos 3 \omega_0 t'-  \cos \omega_0 t') d t'+\frac{1}{8  \omega_0^4} \int\limits_{-\infty}^t 
 \left[\frac{\pa}{\pa t'} \tilde{\hat{W}}^3(t') \right]
 \tilde{\hat{W}}(t') \tilde{\Psi}(t')  ( \frac{1}{9}  \cos\omega_0 t'\cos 3 \omega_0 t'-  \cos^2 \omega_0 t') d t' \nonumber \\
&+\frac{1}{48 \omega_0^4} \int\limits_{-\infty}^t 
\left[\frac{\pa}{\pa t'} \tilde{\hat{W}}^4(t') \right]
 \tilde{\Psi}(t')  ( \frac{1}{4}  \cos 4 \omega_0 t'- \cos 2 \omega_0 t') d t'+\frac{1}{48 i\omega_0^4} \int\limits_{-\infty}^t 
 \tilde{\hat{W}}^5(t') 
 \tilde{\Psi}(t')  \cos\omega_0 t'( \frac{1}{4}  \cos 4 \omega_0 t'- \cos 2 \omega_0 t') d t', \nonumber
\end{align}
% \end{split}
%\end{equation}

% \frac{1}{i} \cos\omega_0 t' \tilde{\hat{W}}(t') \tilde{\Psi}(t')

As all the previous equations starting from Eq.~(\ref{pre}), Eq.~(\ref{ttag}) is exact at any time $t$. Upon the end of the pulse,
$t \to +\infty$, $\hat{W}(t)\to 0$, as do all its time derivatives. Therefore, all the out-of-integrals terms on RHS  of Eq.~(\ref{ttag}), except for $\Psi_0$, (terms from 2nd to 16th)  become zero. We, therefore, can write
\begin{equation}
\begin{split}
&\tilde{ \Psi}(+\infty) = \Psi_0 
 +\frac{1}{4\omega_0^2} \int\limits_{-\infty}^\infty
 \left[ \frac{\pa}{\pa t} \tilde{\hat{W}}(t) ,
 \tilde{\hat{W}}(t) \right]\tilde{\Psi}(t)  d t  -\frac{1}{2 \omega_0^4} \int\limits_{-\infty}^\infty
 \left[ \frac{\pa^3}{\pa t^3} \tilde{\hat{W}}(t) \right]
 \tilde{\hat{W}}(t) \tilde{\Psi}_0    d t\\
 &+\frac{1}{32 \omega_0^4} \int\limits_{-\infty}^\infty
   \left[ \frac{\pa}{\pa t} \tilde{\hat{W}}(t) ,
 \tilde{\hat{W}}(t)\right] 
\tilde{\hat{W}}^2(t)
 \tilde{\Psi}_0 d t 
 +\frac{1}{16 \omega_0^4} \int\limits_{-\infty}^\infty 
  \left[\frac{\pa}{\pa t} \tilde{\hat{W}}^2(t) \right]
 \tilde{\hat{W}}^2(t)
  \tilde{\Psi}_0  d t 
 -\frac{1}{16  \omega_0^4} \int\limits_{-\infty}^\infty
 \left[\frac{\pa}{\pa t} \tilde{\hat{W}}^3(t) \right]
 \tilde{\hat{W}}(t) \tilde{\Psi}_0  d t ,
 \end{split}
\end{equation}
where all the terms of the order $\omega_0^{-n}$, $n>4$, have been neglected, which allowed us to replace $\Psi$ with $\Psi_0$ everywhere but in the 2nd term. The last step is to expand the 2nd term to the same order, which is done by using Eq.~(\ref{ttag}) again
\begin{equation}
\begin{split}
&\tilde{ \Psi}(+\infty) = \Psi_0 +\frac{1}{4\omega_0^2} \int\limits_{-\infty}^\infty
 \left[ \frac{\pa}{\pa t} \tilde{\hat{W}}(t) ,
 \tilde{\hat{W}}(t) \right] \left\{  \Psi_0 
  +\frac{1}{4\omega_0^2} \int\limits_{-\infty}^t 
 \left[ \frac{\pa}{\pa t'} \tilde{\hat{W}}(t') ,
 \tilde{\hat{W}}(t') \right]\tilde{\Psi}_0  d t' +\frac{1}{4\omega_0^2} 
   \tilde{\hat{W}}^2(t) \tilde{\Psi}_0\right\} d t \\
 &-\frac{1}{4 \omega_0^4} \int\limits_{-\infty}^\infty
 \left[ \frac{\pa^3}{\pa t^3} \tilde{\hat{W}}(t), \tilde{\hat{W}}(t)  \right]
\tilde{\Psi}_0    d t 
 -\frac{1}{16  \omega_0^4} \int\limits_{-\infty}^\infty
 \left[\frac{\pa}{\pa t} \tilde{\hat{W}}(t),\tilde{\hat{W}}^3(t)  \right]
  \tilde{\Psi}_0  d t 
 -\frac{3}{32 \omega_0^4} \int\limits_{-\infty}^\infty
  \left[ \frac{\pa}{\pa t} \tilde{\hat{W}}(t) ,
 \tilde{\hat{W}}(t) \right]   
 \tilde{\hat{W}}^2(t)
   \tilde{\Psi}_0  d t
   \\
 &+\frac{1}{32 \omega_0^4} \int\limits_{-\infty}^\infty 
  \left[\frac{\pa}{\pa t} \tilde{\hat{W}}^2(t), \tilde{\hat{W}}^2(t) \right]
  \tilde{\Psi}_0  d t
 +\frac{1}{16  \omega_0^4} \int\limits_{-\infty}^\infty
  \left[\frac{\pa}{\pa t} \tilde{\hat{W}}(t) \right] \tilde{\hat{W}}^3(t)
  \tilde{\Psi}_0  d t ,
 \end{split}
\end{equation}
or
\begin{equation}
\begin{split}
&\tilde{ \Psi}(+\infty) = \Psi_0 +\frac{1}{4\omega_0^2} \int\limits_{-\infty}^\infty
 \left[ \frac{\pa}{\pa t} \tilde{\hat{W}}(t) ,
 \tilde{\hat{W}}(t) \right] \left\{  \Psi_0 
  +\frac{1}{4\omega_0^2} \int\limits_{-\infty}^t 
 \left[ \frac{\pa}{\pa t'} \tilde{\hat{W}}(t') ,
 \tilde{\hat{W}}(t') \right]\tilde{\Psi}_0  d t' \right\} d t \\
 &-\frac{1}{4 \omega_0^4} \int\limits_{-\infty}^\infty
 \left[ \frac{\pa^3}{\pa t^3} \tilde{\hat{W}}(t), \tilde{\hat{W}}(t)  \right]
\tilde{\Psi}_0    d t 
 -\frac{1}{16  \omega_0^4} \int\limits_{-\infty}^\infty
 \left[\frac{\pa}{\pa t} \tilde{\hat{W}}(t),\tilde{\hat{W}}^3(t)  \right]
  \tilde{\Psi}_0  d t 
 -\frac{1}{32 \omega_0^4} \int\limits_{-\infty}^\infty
  \left[ \frac{\pa}{\pa t} \tilde{\hat{W}}(t) ,
 \tilde{\hat{W}}(t) \right]   
 \tilde{\hat{W}}^2(t)
   \tilde{\Psi}_0  d t
   \\
 &+\frac{1}{32 \omega_0^4} \int\limits_{-\infty}^\infty 
  \left[\frac{\pa}{\pa t} \tilde{\hat{W}}^2(t), \tilde{\hat{W}}^2(t) \right]
  \tilde{\Psi}_0  d t
 +\frac{1}{16  \omega_0^4} \int\limits_{-\infty}^\infty
  \left[\frac{\pa}{\pa t} \tilde{\hat{W}}(t) \right] \tilde{\hat{W}}^3(t)
  \tilde{\Psi}_0  d t .
 \end{split}
 \label{AL0}
\end{equation}

Equation (\ref{AL}) follows from Eq.~(\ref{AL0}) after regrouping of the terms.

\end{document}